\documentclass[prd,twocolumn,showpacs,floatfix,amsmath,nofootinbib,amssymb,floatfix]{revtex4}
\usepackage{graphicx,color,dcolumn,booktabs,bm}
\usepackage{longtable,lscape}
\usepackage{txfonts}
\usepackage{overpic}
\usepackage{amssymb}
\usepackage{indentfirst}
\usepackage{feynmf}   %{feynmp}
\usepackage{slashed}  %for Feynman symbols
\usepackage{cases}
\usepackage{color}
\usepackage{multirow}
\usepackage{epstopdf}
\usepackage{graphicx,color,dcolumn,booktabs,bm}
\usepackage[colorlinks,
            citecolor=blue,
            anchorcolor=red,
            menucolor=red,
            linkcolor=red,
            filecolor=red,
            runcolor=red,
            urlcolor=blue,
            frenchlinks=red]{hyperref}

\begin{document}
\title{A systematic study of mass spectra and strong decay of strange mesons}
\author{Cheng-Qun Pang$^{1,2}$}
\author{Jun-Zhang Wang$^{1,3}$}
\author{Xiang Liu$^{1,3}$\footnote{Corresponding author}}
\email{xiangliu@lzu.edu.cn}
\author{Takayuki Matsuki$^{4,5}$}
\email{matsuki@tokyo-kasei.ac.jp}
\affiliation{
$^1$School of Physical Science and Technology, Lanzhou University, Lanzhou 730000, China\\
$^2$College of Physics and Electronic Information Engineering, Qinghai Normal University, Xining, 810000, China\\
$^3$Research Center for Hadron and CSR Physics, Lanzhou University and Institute of Modern Physics of CAS, Lanzhou 730000, China\\
$^4$Tokyo Kasei University, 1-18-1 Kaga, Itabashi, Tokyo 173-8602, Japan\\
$^5$Theoretical Research Division, Nishina Center, RIKEN, Saitama 351-0198, Japan}

\vspace{2cm}
\begin{abstract}

The mass spectrum of the kaon family is analyzed by the {modified} Godfrey-Isgur model with a color screening effect approximating the kaon as a heavy-light meson system. This analysis gives us the structure and possible assignments of the observed kaon candidates, which can be tested by comparing the theoretical results of their two-body strong decays with the experimental data. Additionally, prediction of some partial decay widths is made on the kaons still missing in experiment. This study is crucial to establishing the kaon family and searching for their higher excitations in the future.

%In this paper, we pay attention to kaon family. Since kaon contains a strange and a $\bar{u}/\bar{d}$ quarks, kaon can be approximately treated as heavy-light meson system. In this work, we adopt modified Godfrey-Isgur quark model by including color screening effect to give the analysis of mass spectrum of kaon family, by which we can obtain the structure information of these observed kaon candidates. And then, we further test these possible assignments by comparing the theoretical result of their two-body strong decays with the experimental data. Additionally, we also predict the behaviors of some partial decay widths of these kaons, which are still missing in experiment. This study is crucial to establish kaon family and future search for their higher excitations.
\end{abstract}

\maketitle

\section{Introduction}

As an important part of the meson family, the kaon sub-family has become more and more abundant with experimental progress on the observations of kaons in the past decades. Until now, Particle Data Group (PDG) has collected dozens of kaons \cite{Agashe:2014kda}.
%In Table \ref{list}, we briefly review these observed kaon mesons listed in the PDG.
When facing so abundant {kaons}, it is one of the main tasks of the present study of light hadron spectroscopy how to categorize them into the  family and another task is to investigate its higher radial and orbital excitations.

Before the present work, there were some theoretical papers related to kaons. For example,
thirty years ago, Godfrey and Isgur \cite{Godfrey:1985xj} developed a relativistic quark model, the so-called Godfrey-Isgur (GI) model, by which they studied the mass spectrum of hadrons including kaons. In 2002, Barnes {\it et al.} \cite{Barnes:2002mu} further investigated the strong decays of the observed kaons, which have masses less than 2.2 GeV, where
the $^3P_0$ quark model associated with a simple harmonic oscillator (SHO) wave function was adopted in their calculation. In 2009, Ebert {\it et al.} \cite{Ebert:2009ub} analyzed the mass spectrum and Regge trajectories of kaons by their relativistic quark model.

Due to the present experimental progress on kaons, it is a suitable time to systematically carry out phenomenological study of kaons. In this work, we first calculate the mass spectra of the kaon family by applying the modified GI model \cite{Song:2015nia,Song:2015fha},
where the screening effect is taken into account. Fitting some well established kaon states, we fix the parameters in the model, which are adopted when calculating the masses of other kaon states. Comparing theoretical results with experimental data, we obtain the structure information of the discussed kaons.  Especially, we predict some radial ground  states of kaon which are still missing in experiments, e.g. $K_4(2310)(1^1G_4)$. Using our potential model  approach, the spatial wave functions of the kaons studied can be numerically calculated, which we take as input when studying their Okubo-Zweig-Iizuka (OZI)-allowed two-body strong decays.
For further testing the properties of the kaons, we study their OZI-allowed two-body strong decays, which provide valuable information of their partial and total decay widths, where one uses the quark pair creation (QPC) model which was proposed in Ref. \cite{Micu:1968mk} and extensively applied to studies of other hadrons in Refs. \cite{LeYaouanc:1972ae,
vanBeveren:1979bd,vanBeveren:1982qb,LeYaouanc:1988fx,roberts,Capstick:1993kb,Blundell:1995ev,Ackleh:1996yt,Capstick:1996ib,Bonnaz:2001aj,
Close:2005se,Zhang:2006yj,Lu:2006ry,Sun:2009tg,Liu:2009fe,Sun:2010pg,Rijken:2010zza,Yu:2011ta,Zhou:2011sp,Ye:2012gu,Wang:2012wa,
Sun:2013qca,He:2013ttg,Sun:2014wea,Pang:2014laa,Wang:2014sea,Chen:2015iqa}. Analyzing mass spectra and calculating strong decay behaviors, we finally identify their $n^{2S+1}L_J$ quantum numbers, which reflect the inner structure of the kaons under discussion.  Here, we predict the strong decay behaviors of some kaon states, e.g. $K_4(2310)(1^1G_4)$ has a wide width about 710-880 MeV, and mainly decays into $K_4^*{(2045)}\pi $,
 $K_3^*{(1780)}\pi$, $K \rho _3{(1690)} $ and $K a_2$. The study presented in this work is helpful for establishing the kaon family by including more higher radial and orbital excitations.

This paper is organized as follows. After Introduction,
in Sect. \ref{sec2} we  explain the modified Godfrey-Isgur model and the QPC model.
In Sect.
\ref{sec3}, we adopt the modified Godfrey-Isgur model by including the screening effect
to study the mass spectra of the kaon family. Making a comparison between theoretical and experimental results,
we further obtain the structure information of the observed {kaons}.
In Sect. \ref{sec4}, we present the detailed study of the OZI-allowed two-body strong decays of the discussed kaons.
The paper ends with
conclusions and discussion.

\section{Phenomenological quark models adopted in this work}
\label{sec2}

In our calculation, two phenomenological quark models are adopted, i.e., the modified GI model with the color screening effect\footnote{When studying the mass spectrum of mesons, there are approaches like Dyson-Schwinger and Bethe-Salpeter equations, which are directly related to QCD. However, such theory-based or theory-linked approaches still have some limitations for describing higher excitations of mesons. Instead of theoretical approaches, one may apply phenomenological models to deal with such subjects. Here, the modified GI model is adopted to calculate the mass spectrum of pseudoscalar mesons. }, and the QPC model.
The modified GI quark model is applied to calculate the mass spectrum of the kaon family, by which we obtain the structure information of the observed kaon candidates.
Then, we further test the possible assignments by comparing the theoretical results of their two-body OZI-allowed decays with the experimental data, where the QPC model is used to calculate their strong decays.

In the following, we will introduce these two models.

\subsection{The modified GI model}

First, we introduce the Godfrey-Isgur (GI) relativized quark model and discuss how the GI model is modified by including the color screening effect.
Below we describe the detailed procedure and equations actually done by us because those are necessary in our work but are not familiar to general readers. Some are common to Godfrey and Isgur.

The interaction between quark and antiquark in the GI model \cite{Godfrey:1985xj} is described by the Hamiltonian
\begin{eqnarray}
\tilde{H}=\left(p^2+m_{u/d}^2\right)^{1/2}+ \left(p^2+m_s^2\right)^{1/2} +\tilde{V}_{\mathrm{eff}}\left(\textbf{p},\textbf{r}\right), \label{Eq:Htot}
\end{eqnarray}
where $m_{u/d}$ and $m_{s}$ are the masses of $u/d$ and $s$ quarks, respectively, i.e., $m_{u}=m_d=220$ MeV, $m_s=419 \text{ MeV}$. $\tilde{V}_{\mathrm{eff}}(\textbf{p},\textbf{r}) = \tilde{H}^{\mathrm{conf}}+\tilde{H}^{\mathrm{hyp}}+\tilde{H}^{\mathrm{SO}}\label{1} $ is the effective potential of the $q\bar{q}$ interaction which can be obtained from on-shell $q\bar{q}$ scattering amplitudes in the center-of-mass (CM) frame \cite{Godfrey:1985xj} and relativistic effect corrections. The quantities with tilde will be defined later. On the other hand
$\tilde{V}_{\mathrm{eff}}(\textbf{p},\textbf{r})$ also consists of two main parts. The first one is a $\gamma^{\mu}\otimes\gamma_{\mu}$ short-distance interaction of one-gluon-exchange and the second part is a $1\otimes1$ long-distance color confining linear interaction which is suggested by the lattice QCD \cite{lang1982potential,booth1992running,bali2001qcd,Kawanai2011Interquark,kawanai2012interquark}.

In the nonrelativistic limit,
$V_{\mathrm{eff}}(\textbf{p},\textbf{r})$ without tilde is transformed into the familiar nonrelativistic potential $V_{\mathrm{eff}}(r)$ \cite{Godfrey:1985xj,Lucha:1991vn}
\begin{eqnarray}
V_{\mathrm{eff}}(r)=H^{\mathrm{conf}}+H^{\mathrm{hyp}}+H^{\mathrm{so}}\label{1}
\end{eqnarray}
with
\begin{align}
H^{\mathrm{conf}}&=\Big[-\frac{3}{4}(c+br)+\frac{\alpha_s(r)}{r}\Big](\bm{F}_1\cdot\bm{F}_2) \nonumber \\
 &=S(r)+G(r) , \label{3}\\
H^{\mathrm{hyp}}&=-\frac{\alpha_s(r)}{m_{u/d}m_{s}}\Bigg[\frac{8\pi}{3}\bm{S}_1\cdot\bm{S}_2\delta^3 (\bm r) +\frac{1}{r^3}\Big(\frac{3\bm{S}_1\cdot\bm r \bm{S}_2\cdot\bm r}{r^2} \nonumber  \\ \label{3.1}
&\quad -\bm{S}_1\cdot\bm{S}_2\Big)\Bigg] (\bm{F}_1\cdot\bm{F}_2),  \\
H^{\mathrm{so}}=&H^{\mathrm{so(cm)}}+H^{\mathrm{so(tp)}},  \label{3.2}
\end{align}
where $H^{\mathrm{conf}}$ is the spin-independent potential which contains a linear confining
potential  $S(r)=b r+c$  and the one-gluon exchange potential $G(r)=-4\alpha_s(r)/3r$, $H^{\mathrm{hyp}}$ and $H^{\mathrm{SO}}$ are the color-hyperfine interaction and the spin-orbit interaction, respectively. It can be noted that $\bm{F}_1(\bm{F}_2)=\bm{\lambda}_1(-\bm{\lambda}^*_2)/2$, where $\bm{\lambda}_i$ is Gell-Mann matrix. For the meson, $(\bm{F}_1\cdot\bm{F}_2)=-4/3$. Additionally, the subscripts 1 and 2 denote quark
and antiquark, respectively.

 In Eqs. (\ref{3}) and (\ref{3.1}), the running coupling constant $\alpha_s(r)$ has following form
 \begin{align}
  \alpha_s(r)=\sum_{k}\frac{2\alpha_k}{\sqrt{\pi}} \int_{0}^{\gamma_k r}e^{-x^2}dx,
 \end{align}
 where $k$ is from 1 to 3 and corresponding $\alpha_k$ and $\gamma_k$ are constant, $\alpha_{1,2,3}=0.25,0.15,0.2$  and $\gamma_{1,2,3}=\frac{1}{2},\frac{\sqrt{10}}{2},\frac{\sqrt{1000}}{2}$ \cite{Godfrey:1985xj}. For the color-hyperfine interaction $H^{\mathrm{hyp}}$, the first term stands for contact interaction and second term is a typical form of tensor interaction, here  $\bm{S}_1(\bm{S}_2)$ denotes the spin of the quark (antiquark).

In Eq. (\ref{3.2}), the spin-orbit interaction can be divided into two types in which $H^{\mathrm{so(cm)}}$ is the
color-magnetic term and $H^{\mathrm{so(tp)}}$ is the Thomas-precession term. Their expression can be written as
\begin{eqnarray}
H^{\mathrm{so(cm)}}=-\frac{\alpha_s(r)}{r^3}\left(\frac{1}{m_{u/d}}+\frac{1}{m_{s}}\right)\left(\frac{\bm{S}_1}{m_{u/d}}+\frac{\bm{S}_2}{m_{s}}\right)\cdot
\bm{L}(\bm{F}_1\cdot\bm{F}_2),
\end{eqnarray}
\begin{eqnarray}
H^{\mathrm{so(tp)}}=-\frac{1}{2r}\frac{\partial H^{\mathrm{conf}}}{\partial
r}\Bigg(\frac{\bm{S}_1}{m^2_{u/d}}+\frac{\bm{S}_2}{m^2_{s}}\Bigg)\cdot \bm{L},
\end{eqnarray}
where $\bm{L}$ is the orbital momentum between quark and antiquark.

Noting that the above interaction potentials are obtained in the nonrelativistic limit, and they can optimized by introducing the phenomenological relativistic effects.  In the GI model, the relativistic effects are imposed into the model mainly by two ways. Firstly, a smearing function  $\rho(\bm{r-r}')$ is introduced to incorporate the effects of an internal motion inside a meson and nonlocality of interactions between quark and antiquark. A smearing transformation is given by
\begin{equation}\label{smear}
\tilde{f}(r)=\int d^3r'\rho(r-r')f(r'),
\end{equation}
with
\begin{equation}
\rho \left(r-r'\right)=\frac{\sigma^3}{\pi ^{3/2}}e^{-\sigma^2}\left(r-r'\right)^2,
\end{equation}

\begin{equation}\label{rhoij}
{
\sigma=\sqrt{s^2 \left(\frac{2 m_{u/d} m_{s}}{m_{u/d}+m_{s}}\right){}^2+\sigma _0^2 \left(\frac{1}{2} \left(\frac{4 m_{u/d} m_{s}}{(m_{u/d}+m_{s})^2}\right)^4+\frac{1}{2}\right)}.}
\end{equation}
\iffalse
\begin{equation}
\tau _{{k12}}=\frac{1}{\sqrt{\frac{1}{\sigma _{{12}}^2}+\frac{1}{\gamma _k^2}}}.
\end{equation}

\begin{eqnarray}
&&\rho_{12}(\bm{r-r}')=\frac{\sigma^3_{12}}{\pi^{3/2}}\text{e}^{-\sigma^2_{12}(\bm{r-r}')^2}, \label{4-1} \\
&&\sigma_{12}^2=\sigma_0^2\Bigg[\frac{1}{2}+\frac{1}{2}\left(\frac{4m_1m_2}{(m_1+m_2)^2}\right)^4\Bigg]+
s^2\left(\frac{2m_1m_2}{m_1+m_2}\right)^2, \nonumber
\end{eqnarray}
\fi
where {$\sigma_0$ =1.80 GeV and $s$=1.55, are the universal parameters in the GI model, $f(r)$ is a arbitrary function and notation tilde stands for that the expression has been performed smearing transformation.} % a parameter $\sigma_{12}$ expresses two types of mesons
%indicates a power of relativity.
%
By smearing transformation, the one-gluon exchange potential $G(r)=-4\alpha_s(r)/(3r)$ and linear confined potential $S(r)=br+c$ are changed as
\begin{equation}
\tilde{G}(r)=-\sum_{k}\frac{8\alpha_k}{3\sqrt{\pi}r}\int_{0}^{\tau_{k}r}e^{-x^2}dx,
\end{equation}
\begin{equation}\label{4-1}
\tilde{S}(r)=br\left[\frac{e^{-\sigma^2r^2}}{\sqrt{\pi}\sigma r}+\left(1+\frac{1}{2\sigma^2r^2}\right)\frac{2}{\sqrt{\pi}}\int_0^{\sigma r}e^{-x^2}dx\right]+c\, ,
\end{equation}
where
\begin{equation}
\tau _{{k}}=\frac{1}{\sqrt{\frac{1}{\sigma^2}+\frac{1}{\gamma _k^2}}}
\end{equation}
\iffalse
It is worth mentioned that if $m_1=m_2$ and both of them are large, and $\sigma_{12}$ also becomes large. In this situations,  $\rho_{12}(\bm{r-r}')\to\delta^3(\bm{r-r}')$, and the second term becomes $(sm_1)^2$, hence this smearing is reasonable for a heavy quarkonium. It is also suitable to the heavy-light system because $\sigma_{12}$ gets its minimum when $m_2 >> m_1$.
\fi
%

%
%

Secondly, a general expression of the potential should be dependent on the CM momentum of the interacting quarks. So the momentum-dependent effect is achieved by introducing momentum-dependent factors which will go to unity in the nonrelativistic limit. In a semiquantitative relativistic treatment, the smeared one-gluon exchange potential term $\tilde{G}(r)$ and the smeared hyperfine interactions (or spin-orbit interaction)  $\tilde{V}^i$  should be modified with following momentum-dependent factors

\begin{equation}\label{G}
{G}^\prime (r)= \left(1+\frac{p^2}{E_1E_2}\right)^{1/2}\tilde{G}(r)\left(1 +\frac{p^2}{E_1E_2}\right)^{1/2},
\end{equation}
\begin{equation}
\tilde{G}^{so(v)}_{ij}=\left(\frac{m_im_j}{E_iE_j}\right)^{1/2+\epsilon_{so(v)}} \tilde{G}(r)\left(\frac{m_im_j}{E_iE_j}\right)^{1/2+\epsilon_{so(v)}},
\end{equation}
\begin{equation}
\tilde{G}^{c}_{12}=\left(\frac{m_1m_2}{E_1E_2}\right)^{1/2+\epsilon_{c}} \tilde{G}(r)\left(\frac{m_1m_2}{E_1E_2}\right)^{1/2+\epsilon_{c}},
\end{equation}
\begin{equation}
\tilde{G}^{t}_{12}=\left(\frac{m_1m_2}{E_1E_2}\right)^{1/2+\epsilon_{t}} \tilde{G}(r)\left(\frac{m_1m_2}{E_1E_2}\right)^{1/2+\epsilon_{t}},
\end{equation}
\begin{equation}
\tilde{S}^{so(s)}_{11}=\left(\frac{m_1^2}{E_1^2}\right)^{1/2+\epsilon_{so(s)}} \tilde{S}(r)\left(\frac{m_1^2}{E_1^2}\right)^{1/2+\epsilon_{so(s)}},
\end{equation}
\begin{equation}
\tilde{S}^{so(s)}_{22}=\left(\frac{m_2^2}{E_2^2}\right)^{1/2+\epsilon_{so(s)}} \tilde{S}(r)\left(\frac{m_2^2}{E_2^2}\right)^{1/2+\epsilon_{so(s)}},
\end{equation}
where $E_1=\sqrt{m_{u/d}^2+p^2}$ and $E_2=\sqrt{m_{s}^2+p^2}$ are the energies of the quark and antiquark in the meson,  and $m_1=m_{u/d}$, $m_2=m_{s}$, $\epsilon_i$ is parameter for a different type of hyperfine and spin-orbit interactions, which include the {{contact}}, tensor, vector spin-orbit and scalar spin-orbit potentials. Here, vector spin-orbit and scalar spin-orbit potentials correspond to the Eq. (\ref{3.2}) related to one gluon exchange and confinement term, respectively.
So the total Hamiltonian can be written as
\begin{eqnarray}
\hat{H}=\left(p^2+m_{u/d}^2\right)^{1/2}+ \left(p^2+m_{s}^2\right)^{1/2} +\tilde{H}^{conf}+\tilde{H}^{hyp}+\tilde{H}^{so}
\label{Eq:Htot1}
\end{eqnarray}
with
\begin{equation}
\tilde{H}_{12}^{conf}={G}^\prime (r)+\tilde{S}(r), \\
\end{equation}

\begin{equation}
\tilde{H}^{so}=\tilde{H}^{so(v)}+\tilde{H}^{so(s)},
\end{equation}
where
\begin{eqnarray}
\tilde{H}^{so(v)}&=&\frac{\textbf{S}_1 \cdot\textbf{L} }{2m_{u/d}^2r}\frac{\partial\tilde{G}_{11}^{so(v)}}{\partial r}+\frac{\textbf{S}_2 \cdot\textbf{L}}{2m_s^2r}\frac{\partial\tilde{G}_{22}^{so(v)}}{\partial r} \nonumber \\
&&+\frac{(\textbf{S}_1+\textbf{S}_2) \cdot\textbf{L} }{m_{u/d}m_s}\frac{1}{r}\frac{\partial \tilde{G}_{12}^{so(v)}}{\partial r},
\end{eqnarray}

\begin{eqnarray}
\tilde{H}^{so(s)}=-\frac{\textbf{S}_1 \cdot\textbf{L}}{2m_{u/d}^2r}\frac{\partial\tilde{S}_{11}^{so(s)}}{\partial r}-\frac{\textbf{S}_2 \cdot\textbf{L}}{2m_s^2r}\frac{\partial\tilde{S}_{22}^{so(s)}}{\partial r},
\end{eqnarray}

\begin{equation}
\tilde{H}_{12}^{hyp}=\tilde{H}^{tensor}_{12}+\tilde{H}^{c}_{12},
\end{equation}
where
\begin{equation}
\tilde{H}_{12}^{tensor}=-\left(\frac{\textbf{S}_1\cdot \textbf{r} \textbf{S}_2\cdot \textbf{r}/r^2-\frac{1}{3}\textbf{S}_1\cdot \textbf{S}_2}{m_{u/d}m_{s}}\right)\left(\frac{\partial^2}{\partial r^2}-\frac{1}{r}\frac{\partial}{\partial r}\right)\tilde{G}^t_{12},\nonumber
\end{equation}

\begin{equation}
\tilde{H}^{c}_{12}=\frac{2\textbf{S}_1\cdot \textbf{S}_2}{3m_{u/d}m_{s}}\nabla^2\tilde{G}^c_{12}.\nonumber
\end{equation}

For solving Schr$\ddot{\text{o}}$dinger equation $\hat{H}\Psi=E\Psi$ with $\hat{H}$ shown in Eq. (\ref{Eq:Htot1}), the simple harmonic {oscillators (SHO)} wave function will be employed. In the configuration space, SHO wave function has the form
\begin{align}
\Psi_{nLM_L}(\mathbf{r})=R_{nL}(r, \beta)Y_{LM_L}(\Omega_r),\nonumber\\
\Psi_{nLM_L}(\mathbf{p})=R_{nL}(p, \beta)Y_{LM_L}(\Omega_p),
\end{align}
with
\begin{eqnarray}
&R_{nL}(r,\beta)=\beta^{3/2}\sqrt{\frac{2n!}{\Gamma(n+L+3/2)}}(\beta r)^{L}
 e^{\frac{-r^2 \beta^2}{2}}L_{n}^{L+1/2}(\beta^2r^2),
 \nonumber\\
 &R_{nL}(p,\beta)=\frac{(-1)^n(-i)^L}{ \beta ^{3/2}}e^{-\frac{p^2}{2 \beta ^2}}\sqrt{\frac{2n!}{\Gamma(n+L+3/2)}}{(\frac{p}{\beta})}^{L}
 L_{n}^{L+1/2}(\frac{p^2}{ \beta ^2}),
\end{eqnarray}
where $Y_{LM_L}(\Omega)$ is spherical harmonic function with {{orbital}} angular momentum quantum number L, and $L_{n-1}^{L+1/2}(x)$ is an associated Laguerre polynomial, and $\beta$ is a parameter of oscillator radial wave function. A series of SHO wave function with different radial quantum number $n$ can be regarded as a complete basis to expand the exact radial wave function of meson state, in this case, the meson mass spectrum can be obtained by diagonalizing the Hamiltonian matrix of Eq. (\ref{Eq:Htot1}) based the above SHO basis. The total wave function of meson is composed by color, flavor, spin, space wave function, and the spin wave functions $\chi$ are
\begin{align}
&\chi_{00}=\frac{1}{\sqrt{2}}(\uparrow\downarrow-\downarrow\uparrow), \nonumber \\
&\chi_{11}=\uparrow\uparrow , \nonumber \\
&\chi_{10}=\frac{1}{\sqrt{2}}(\uparrow\downarrow+\downarrow\uparrow), \nonumber \\
&\chi_{1-1}=\downarrow\downarrow.
\end{align}
The space-spin wave function $R_{nL}(r, \beta)\phi_{LSJM}$ with total angular quantum number {\it J} can be constructed by coupling $L\otimes S$ and has form
\begin{align}
\phi_{LSJM}=\sum_{M_LM_S}C(L M_L S M_S; JM)Y_{LM_L}(\Omega_r)\chi_{SM_S},
\end{align}
where $C(L M_L S M_S; JM)$ is Clebsch-Gordan coefficient. {For the matrix element $\langle \alpha | \hat{V}(r,\hat{p}) | \beta \rangle$ where $| \alpha \rangle$ and $| \beta \rangle$ are arbitrary SHO basis with quantum number $\{n,J,L,S\}$ and $\{n^{\prime},J^{\prime},L^{\prime},S^{\prime}\}$}. It is noted that the color and flavor wave function of meson have no contributions for the matrix element of Hamiltonian, and there are general expression
\begin{align}
&\langle \alpha | \hat{V}(r,\hat{p}) | \beta \rangle \nonumber \\
&={\langle \alpha | f(p)g(r) | \beta \rangle} \nonumber \\
&=\sum_{n}{\langle \alpha | f(p) |  n\rangle   \langle n |   g(r) | \beta \rangle}.
\end{align}
After calculating each matrix element, the mass and wave function of meson could be obtained and they also are available to the following strong decay process.

Although the GI model has achieved great success in describing the meson spectrum, there still exists a discrepancy between the predictions
 given by  the GI model and recent experimental observations. The previous work \cite{Song:2015nia} presents a modified GI model with a screening potential whose predictions can be well consistent
 with the experiment data for the charm-strange mesons. For higher excitation states, the authors of Ref. \cite{Song:2015nia} believe that a {{screening}} effect plays a very important role which could be introduced by the transformation
 $br+c\rightarrow \frac{b(1-e^{-\mu r})}{\mu}+c$, {where $\mu$ is} a screening parameter whose particular value is need to be fixed by the comparisons between theory and experiment.
Modified confinement potential also need to make similar relativistic correction which has been mentioned in  the GI model.
 Then, we further write $V^{\mathrm{scr}}(r)$ as the way given in Eq. (\ref{4-1}),
\begin{eqnarray}
\tilde V^{\mathrm{scr}}(r)&=& \int d^3 \bm{r}^\prime
\rho (\bm{r-r^\prime})\frac{b(1-e^{-\mu r'})}{\mu}.\label{5}
\end{eqnarray}
By inserting the form of  $\rho(\bm{r-r^\prime})$  in Eq.~(\ref{rhoij}) into the above expression and finishing this integration, the concrete expression for $\tilde V^{\mathrm{scr}}(r)$ is given by
\begin{eqnarray}
\tilde V^{\mathrm{scr}}(r)&=& \frac{b}{\mu r}\Bigg[r+e^{\frac{\mu^2}{4 \sigma^2}+\mu r}\frac{\mu+2r\sigma^2}{2\sigma^2}\Bigg(\frac{1}{\sqrt{\pi}}
\int_0^{\frac{\mu+2r\sigma^2}{2\sigma}}e^{-x^2}dx-\frac{1}{2}\Bigg) \nonumber\\
&&-e^{\frac{\mu^2}{4 \sigma^2}-\mu r}\frac{\mu-2r\sigma^2}{2\sigma^2}\Bigg(\frac{1}{\sqrt{\pi}}
\int_0^{\frac{\mu-2r\sigma^2}{2\sigma}}e^{-x^2}dx-\frac{1}{2}\Bigg)\Bigg]. \label{Eq:pot}
\end{eqnarray}
It is worth mentioning that after the confinement potential is replaced with a screening potential, other treatments are similar to the original GI model including the calculation of matrix elements of the Hamiltonian.
\subsection{The QPC model}

The QPC model was first proposed by Micu \cite{Micu:1968mk} and further developed by the Orsay group \cite{LeYaouanc:1972ae,LeYaouanc:1973xz,LeYaouanc:1974mr,LeYaouanc:1977gm,LeYaouanc:1977ux}. This model was
widely applied to the OZI-allowed two-body strong decay of hadrons
in Refs. \cite{vanBeveren:1979bd,vanBeveren:1982qb,Capstick:1993kb,Page:1995rh,Titov:1995si,Ackleh:1996yt,Blundell:1996as,
Bonnaz:2001aj,Zhou:2004mw,Lu:2006ry,Zhang:2006yj,Luo:2009wu,Sun:2009tg,Liu:2009fe,Sun:2010pg,Rijken:2010zza,Ye:2012gu,
Wang:2012wa,He:2013ttg,Sun:2013qca,Pang:2014laa,Wang:2014sea}.

For a decay process $A\to B+C$, we can write
\begin{eqnarray}
\langle BC|\mathcal{T}|A \rangle = \delta ^3(\mathbf{P}_B+\mathbf{P}_C)\mathcal{M}^{{M}_{J_{A}}M_{J_{B}}M_{J_{C}}},
\end{eqnarray}
where $\mathbf{P}_{B(C)}$ is a three-momentum of a meson $B(C)$ in the rest frame of a meson $A$. A superscript $M_{J_{i}}\, (i=A,B,C)$ denotes an orbital
magnetic momentum. The transition operator $\mathcal{T}$ is introduced to describe a quark-antiquark pair creation from vacuum, which has the quantum number
$J^{PC}=0^{++}$, i.e., $\mathcal{T}$ can be expressed as
\begin{eqnarray}
\mathcal{T}& = &-3\gamma \sum_{m}\langle 1m;1~-m|00\rangle\int d \mathbf{p}_3d\mathbf{p}_4\delta ^3 (\mathbf{p}_3+\mathbf{p}_4) \nonumber \\
 && \times \mathcal{Y}_{1m}\left(\frac{\textbf{p}_3-\mathbf{p}_4}{2}\right)\chi _{1,-m}^{34}\phi _{0}^{34}
\left(\omega_{0}^{34}\right)_{ij}b_{3i}^{\dag}(\mathbf{p}_3)d_{4j}^{\dag}(\mathbf{p}_4),
\end{eqnarray}
which is constructed in a completely phenomenological way to reflect the creation of a quark-antiquark pair from vacuum, where the quark and antiquark are
denoted by indices $3$ and $4$, respectively.
A dimensionless parameter $\gamma$ depicts the strength of the creation of $q\bar{q}$ from vacuum, where the concrete values of the parameter $R$ which will be discussed in the later section. $\mathcal{Y}_{\ell m}(\mathbf{p})={|\mathbf{p}|^{\ell}}Y_{\ell
m}(\mathbf{p})$ are the solid harmonics. $\chi$, $\phi$, and $\omega$ denote the spin, flavor, and color wave functions respectively, which can be treated separately.
Subindices $i$ and $j$ denote the color of a $q\bar{q}$ pair.

By the Jacob-Wick formula \cite{Jacob:1959at}, the decay amplitude is expressed as
\begin{eqnarray}
\mathcal{M}^{JL}(\mathbf{P})&=&\frac{\sqrt{4\pi(2L+1)}}{2J_A+1}\sum_{M_{J_B}M_{J_C}}\langle L0;JM_{J_A}|J_AM_{J_A}\rangle \nonumber \\
&&\times \langle J_BM_{J_B};J_CM_{J_C}|{J_A}M_{J_A}\rangle \mathcal{M}^{M_{J_{A}}M_{J_B}M_{J_C}},
\end{eqnarray}
and the general decay width reads
\begin{eqnarray}
\Gamma&=&\frac{\pi}{4} \frac{|\mathbf{P}|}{m_A^2}\sum_{J,L}|\mathcal{M}^{JL}(\mathbf{P})|^2,
\end{eqnarray}
where $m_{A}$ is the mass of an initial state $A$.
In our calculation, we need the spatial wave functions of the discussed kaons and iso-scalar and iso-vector light mesons.
which can be numerically obtained by the modified GI model.

\section{Mass spectrum analysis}\label{sec3}

%\subsection{Parameters of the modified GI model}

Although the GI model has succeeded in describing the ground states of the kaon family, it does not well describe the excited states. Since unquenched effects are important for a heavy-light system, it is better to adopt the modified GI model ({MGI}) \cite{Song:2015nia,Song:2015fha} which uses a screening potential with a new parameter $\mu$. The parameter $\mu$ describes inverse of the size of screening. To use the {MGI} model to calculate the kaon family spectra, it is better to determine the value of a new parameter $\mu$ considering two features:
The first is when we use the same parameter set as in Ref. \cite{Godfrey:1985xj} and add a new parameter $\mu$, the mass of the ground state of the kaon family will be lower than the experiments. The second one is  the value of  $\mu$ may be not so small like the one in Refs. \cite{Song:2015nia,Song:2015fha}. In fact, in bottomonium and charmonium states, Refs. \cite{Li:2009nr,Li:2009zu} give  the $\mu$ value about 0.1 GeV which is larger than  the one in Refs. \cite{Song:2015nia,Song:2015fha}. Since we do not know the real value of $\mu$ in the kaon family beforehand, we need to adjust the parameters by fitting with the experiments data.
At first, the quark masses should be the same for all meson families.
% So in our fit, the parameter of the quark mass is not included.
Secondly, we do not adjust the values of $\Lambda$ and $\alpha_s$ for the same reason.
Since $\sigma_0$ and $s$ are universal parameters which are resolved by the $Q\bar{Q}$ system, we do not vary them in our fit. The confining term $br+c$ will be replaced by the screening potential, so their parameters should be fitted again. The relativistic effects should be adapted to a different system with the different quark masses. So we fix the following seven parameters listed in Table \ref{SGIfit1} by fitting eleven experimental data which is listed in Table \ref{fitstate}.
\renewcommand{\arraystretch}{1.2}
\begin{table}[htbp]
\caption{Parameters and their values in this work and GI models. \label{SGIfit1}}
\begin{center}
\begin{tabular}{cccc}
\toprule[1pt]\toprule[1pt]
Parameter &  This work &GI \cite{Godfrey:1985xj} \\
 \midrule[1pt]
$b$ &0.2555&0.18\\
$c$ &-0.3492&-0.253\\
$\mu$&0.1 &0 &  \\
$\epsilon_{sov}$&-0.01700&-0.035\\
$\epsilon_c$&-0.1396& -0.168\\
$\epsilon_t$&0.03600& 0.025\\
$\epsilon_{sos}$&0.06772&0.055\\
\bottomrule[1pt]\bottomrule[1pt]
\end{tabular}
\end{center}
\end{table}
%\cleardoublepage
%\newpage

 \renewcommand{\arraystretch}{1.2}
\begin{table}[htbp]
\caption{The experimental data~\cite{Agashe:2014kda} fitted in our work. $\chi^2=({\rm {(Th-Exp)}/{Error}})^2$, where Th, Exp, and Error represent the theoretical, experimental results, and experimental error, respectively, and $n$ is the number of the experiment data.  We select some established kaon states in PDG \cite{Agashe:2014kda} for our fitting.
% and so as to the later.
The unit of the mass is MeV. \label{fitstate}}
\begin{center}
\begin{tabular}{cccccccc}
\hline\hline
States&$n^{2S+1}L_J$&This work&GI \cite{Godfrey:1985xj} &Experiment \cite{Agashe:2014kda}&Error in fitting  \\
\hline
\hline
 $K$ &$1^1 {S}_0$ &  497.7 & 461.5 & ${497.6\pm0.013}$&1.3\\
 $K^*(892)$    &$1^3 {S}_1$ &  896 & 902.8 & ${895.8\pm0.19}$ &1.9\\
 $K_0^*(1430)$  &$1^3 {P}_0$ &  1257  & 1234 & ${1425\pm50}$&50\\
 $K_2^*(1430)$    &$1^3 {P}_2$ &  1431 & 1428 & ${1432.4\pm1.3}$ &1.3\\
 $K^*(1680)$    &$1^3 {D}_1 $&  1766  & 1776 & ${1717\pm27}$ &27\\
 $K^*_3(1780)$    &$1^3 {D}_3 $&  1781  & 1794 & ${1776\pm7}$&7\\
 $K_4^*(2045)$    &$1^3 {F}_4 $&  2058  & 2108 & ${2045\pm9}$&9\\
 $K_5^*(2380)$    &$1^3 {G}_5 $&  2286  & 2388 & ${2382\pm14\pm19}$ &24\\
 $K(1460)$    &$2^1 {S}_0$ &  1457  & 1454 & ${1460}$ &20\\
 $K^*(1410)$    &$2^3 {S}_1 $&  1548  & 1579 & ${1414\pm15}$ &15\\
 $K_0^*(1950)$    &$2^3 {P}_0 $&  1829  & 1890  & ${1945\pm10\pm20}$&22 \\
 \hline
${\chi^2}/{ n}$ &   & 12.6&90.2& \\
\hline
\hline
\end{tabular}
\end{center}
\end{table}
{In Table \ref{fitstate}, we select eleven experimental data of kaons listed in PDG
and optimize these kaon masses to
determine seven parameters in Table \ref{SGIfit1}.
This optimization has $\chi^2/n=12.6$ which is smaller than 90.2 for the GI model as shown in Table \ref{fitstate}.
Another reasons why we choose these kaons to fix the parameters in our model is that there does not exist mixture {between} $n^1L_L$ and $n^3L_L$ states for these kaons.  In order to obtain the optimum values of parameters and global and good fit of eleven data, we set ``Error in fitting'' in Table \ref{fitstate} so that the first two experimental data, corresponding to $K$ and $K^*$, have artificial larger error values instead of the real errors in the brackets in the fourth column.}
{The results listed in Table \ref{fitstate} show that the {MGI} model is better than the GI model since the value of ${\chi^2}/{ n}$ of the {MGI} model is about 7 times smaller than that of the GI model and hence it is safely applied to describe the masses of the selected eleven kaons.}

Although the {MGI} model is better than the GI model to depict eleven experimental data,
we need to indicate that there may exist $\sim \mathcal{O}(100 $ MeV) deviation  between experimental and fitting results
for several kaons, which is shown in Table \ref{fitstate}. Such a difference of experimental and theoretical results
may be due to precision of experiment. For example, there is only one experiment \cite{Agashe:2014kda} for $K_5^*(2380)$ and $K_0^*(1950)$. The confirmation to $K_5^*(2380)$ and $K_0^*(1950)$ is still absent. Thus, further experimental measurement of the resonance parameters of these kaons
will be helpful to clarify this difference of experimental and theoretical results.

%\subsection{Spectra of the kaon family}

By using the parameters shown in Table \ref{SGIfit1} as input, we further calculate the masses of other kaons, which are collected in Table \ref{tab:kmassground1}, where we do not consider the mixing of states with $n^1L_L$ and $n^3L_L$. Usually, there exists mixture of the
$n^1L_L$ and $n^3L_L$ states, i.e., \cite{Matsuki:2010zy},
\begin{equation}
\left( \begin{array}{c} |nL\rangle \\|nL^\prime\rangle \end{array} \right) \approx
\left( \begin{array}{cc} \cos{\theta_{nL}} & \sin{\theta_{nL}} \\
                         -\sin{\theta_{nL}} & \cos{\theta_{nL}} \end{array} \right)
\left( \begin{array}{c} |n^1L_L\rangle\\ |n^3L_L\rangle \end{array} \right),
\label{kmixing}
\end{equation}
{where $|nL \rangle$ and $|nL^\prime \rangle$ are two mixing physical states and $\theta_{nL}$ is the corresponding mixing angle. Introducing such mixing states, we find two mass relations $m(nL)<m(n^1L_L)$ and $m(nL^\prime)>m(n^3L_L)$, which can be applied to identify these observed kaons with the same $J^P$ quantum number.} Thus, we need to combine the mass relations and mass spectrum of kaons listed in Table  \ref{tab:kmassground1} with the experimental data to further shed light on the properties of other observed kaons. We conclude that

\iffalse
In Table \ref{fitstate}, one can notice that the fitting results of the SGI model are very close to the experimental data listed in PDG except for $1^3P_0$, $2^3P_0$, $2^3S_1$ and $1^3G_5$ states.
In Table \ref{tab:kmassground1}, we compare our results with the mass spectra obtained by GI \cite{Godfrey:1985xj} and Ref. \cite{Barnes:2002mu}, and
we can roughly conclude  that our results are slightly different from those of Ref. \cite{Godfrey:1985xj}, Ref.~\cite{Ebert:2009ub}, and experiment results, but are very close to the results of Ref.~\cite{Barnes:2002mu}. After the brief global analysis for the kaon mass spectra, we should make use of our theoretical results to investigate underlying property of some discovered particle states which are not well understood now.
\fi

\begin{table}[htbp]
\tabcolsep=1pt
\caption{ The masses of other kaons obtained by the { MGI} model and comparison with those from other potential models. The unit of the mass is MeV.\label{tab:kmassground1}}
\begin{center}
\renewcommand{\arraystretch}{1.2}
\resizebox{!}{7.6cm}{
\begin{tabular}{ccccccccc}
\hline\hline
$n^{2S+1}L_J$&This work & GI~\cite{Godfrey:1985xj}&Ref.~\cite{Ebert:2009ub}&Ref.~\cite{Barnes:2002mu}&
$n^{2S+1}L_J$&This work\\
\hline
 $1^1 {P}_1$ & 1364& 1352 &--& -- &$3^1 {D}_2 $& 2380 \\
 $1^3 {P}_1$& 1377  & 1366&-- &-- &$3^3 {D}_1 $& 2385 \\
 $1^1 {D}_2$& 1778 & 1791 &1709&--&$3^3 {D}_2 $& 2388 \\
 $1^3 {D}_2$& 1789  & 1804&1824 &--&$3^3 {D}_3 $& 2382 \\
 $1^1 {F}_3$& 2075 & 2131 &2009&2050&$3^1 {F}_3 $& 2550 \\
 $1^3 {F}_2$& 2093 & 2151 &1964&2050&$3^3 {F}_2 $& 2560 \\
 $1^3 {F}_3$& 2084  & 2143 &2080&2050 &$3^3 {F}_3 $& 2546  \\
 $1^1 {G}_4$ & 2309& 2422 &2255&--&$3^3 {F}_4 $& 2533 \\
 $1^3 {G}_3$ & 2336  & 2458 &2207&--&$3^1 {G}_4 $& 2673 \\
 $1^3 {G}_4$ & 2317 & 2433 &2285&--& $3^3 {G}_3 $& 2687\\
 $2^1 {P}_1 $& 1840  & 1897 &1757&1850&$3^3 {G}_4 $& 2677  \\
 $2^3 {P}_1 $& 1861 & 1928 &1893&1850 &$3^3 {G}_5 $& 2662\\
 $2^3 {P}_2$ & 1870 & 1938 &1896&1850&$4^1 {S}_0 $& 2248\\
 $2^1 {D}_2$ & 2121  & 2238 &2066&--&$4^3 {S}_1 $& 2287\\
 $2^3 {D}_1$ & 2127  & 2251 &2063&--&$4^1 {P}_1 $& 2422 \\
 $2^3 {D}_2 $& 2131 & 2254 &2163&--& $4^3 {P}_0 $& 2424\\
 $2^3 {D}_3$ & 2121  & 2237 &2182&--& $4^3 {P}_1 $& 2434 \\
 $2^1 {F}_3$ & 2340  & 2524 &2348&--&$4^3 {P}_2 $& 2438  \\
 $2^3 {F}_2$ & 2356  & 2551 &--&--&$4^1 {D}_2 $& 2570 \\
 $2^3 {F}_3 $& 2347  & 2536 &--&--&$4^3 {D}_1 $& 2573\\
 $2^3 {F}_4$ & 2328  & 2504 &2436&--&$4^3 {D}_2 $& 2575  \\
 $2^1 {G}_4$ & 2520 & 2779 &2575&--&$4^3 {D}_3 $& 2571\\
 $2^3 {G}_3 $& 2540 & 2814 &--&--&$4^1 {F}_3 $& 2688 \\
 $2^3 {G}_4$ & 2526  & 2789 &--&--&$4^3 {F}_2 $& 2695 \\
 $2^3 {G}_5 $& 2504  & 2749 &--&--&$4^3 {F}_3 $& 2691\\
 $3^1 {S}_0$ & 1924  &2065&--& 1860& $4^3 {F}_4 $& 2683\\
 $3^3 {S}_1$ & 1983  &2156&1950& {--}&$4^1 {G}_4 $& 2782 \\
 $3^1 {P}_1$ & 2177 &2164&-- &{--}&$4^3 {G}_3 $& 2790 \\
 $3^3 {P}_0 $& 2176 &2160&--& {--}&$4^3 {G}_4 $& 2785 \\
 $3^3 {P}_1 $& 2192  &2200&--& {--}&$4^3 {G}_5$ & 2776 \\
 $3^3 {P}_2$ & 2198 &2206&--& {--}&\\

\hline
\hline
\end{tabular}
}
\end{center}
\end{table}

\begin{enumerate}

\item{Both $K_1(1270)$ with {$M=(1272\pm7)$} MeV \cite{Agashe:2014kda} and $K_1(1400)$ with mass {$M=(1403\pm7)$} MeV \cite{Agashe:2014kda} have $J^P=1^+$ quantum number.
$K_1(1270)$ and  $K_1(1400)$ are the mixture of $1^1P_1$ and $1^3P_1$ states, i.e., $K_1(1270)$ and  $K_1(1400)$ correspond to  $1P$ and $1 P^\prime$  states, respectively. }

\item{$K_2(1770)$ has $J^P=2^-$ and {$M=(1773\pm8)$} MeV \cite{Agashe:2014kda}, while $K_2(1820)$ has $J^P=2^-$ and {$M=(1816\pm13)$} MeV \cite{Agashe:2014kda}.  $K_2(1770)$ and $K_2(1820)$, which correspond to
the   $1D$ and $1D^\prime$ states, respectively, are the mixture of the $1^1D_2$ and $1^3D_2$ states. }

\item{$K_1(1650)$ has $J^P=1^+$ and {$M=(1650\pm50)$} MeV \cite{Agashe:2014kda}. Since the mass of $K_1(1650)$ is smaller than that of the $2^1P_1$ state obtained in Table \ref{tab:kmassground1}, thus we suggest that $K_1(1650)$ can be assigned as a  $2P$ state. There must exists its partner,  $2P^\prime$ state, which is still missing in experiments. }

\item{We suggest that $K(1830)$ is a $3^1S_0$ state. Later, we will test this assignment by studying its decay behavior.}

\item{$K_2^*(1980)$ with $J^P=2^+$ and {$M=(2020\pm20)$} MeV \cite{Tikhomirov:2003gg} is either a $2^3P_2$ state or a $1^3F_2$ state. }

\item{$K_2(2250)$ with $J^P=2^-$ and {$M=(2247\pm17)$} MeV \cite{Agashe:2014kda} is the candidate of  $2D^\prime$, which is the mixture of the $2^1D_2$ and $2^3D_2$ states.}

\item{$K_3(2320)$ has $J^P=3^+$ and {$M=(2324\pm24)$} MeV \cite{Agashe:2014kda}. The possible assignment of $K_3(2320)$ is the  $2F$ state, which is the mixture of states $K(2^1F_3)$ and $K(2^3F_3)$. As the partner of $K_3(2320)$,  $K(2F^\prime)$  is till absent in experiments. In addition, we should mention that the $1F$ and $1F^\prime$ in the kaon family are still missing. }

\item{$K_4(2500)$ with $J^P=4^-$ and {$M=(2490\pm20)$} MeV \cite{Agashe:2014kda} can be a $2G$  state, while its partner $K(2G^{'})$ and two kaons   $K(1G)$ and $K(1G^\prime)$ are still missing in experiments.}
\end{enumerate}

Surely, the above conclusions of possible quantum states are only from the point of mass spectra view. If we want to clearly study particle properties further, we also need to investigate decay behaviors, especially strong decays, and detailed study will be given in the next section.

\section{OZI-allowed two-body strong decays}\label{sec4}
%\subsection{Parameter in the QPC model}
In the previous section,
calculating the spectra of the kaon family, we obtain kaon wave functions, too, at the same time, which can be used in the QPC model to study the strong decay of the kaon family. The parameter $\gamma$ in the QPC model is determined by fitting with the experiment data \cite{Blundell:1996as}. Thus, there is no free parameter in the QPC model. We obtain $\gamma$ =10.5 as shown in Table \ref{QPCfit}.
\begin{table}[htbp]
\renewcommand{\arraystretch}{1.2}
\caption{The parameter~$\gamma$ fitting in the QPC model. %Here, the values in brackets in the second column represent experimental error.
The unit of the width is MeV. \label{QPCfit}}
\begin{center}
\begin{tabular}{ccccccccc}
\hline\hline
  Channels &Experimental data&Numerical result \\
\hline
  $K^*\to K\pi $ & 50.2$\pm5$ & 33.5  \\
 $K_0^*(1430)\to K \pi$  & 267$\pm36$ & 314\\
 $K_2^*(1430)\to K\pi$  & 48.9$\pm1.7$ &51.5  \\
 $K_2^*(1430)\to K^*\pi$  & 24.8$\pm1.7$ &20.4  \\
 $K_2^*(1430)\to K\rho$  & 8.7$\pm0.8$& 6.13  \\
 $K_2^*(1430)\to K\omega$  & 2.9$\pm0.8$ & 1.82  \\
 $K_3^*(1780)\to K\rho$  & 74$\pm10$ & 20.1 \\
 $K_3^*(1780)\to K^*\pi$  & 45$\pm7$ & 28.5 \\
 $K_3^*(1780)\to K\pi$  & 31.7$\pm3.7$ & 38.1 \\
 $K_4^*(2045)\to K\pi$  & 19.6$\pm3.8$ & 21.0 \\
 $K_4^*(2045)\to K\phi$  & 2.8$\pm1.4$ & 3.80 \\

\hline
  &$\chi ^2$ = 6.8,~$\gamma$=10.5 \\

\hline
\hline
\end{tabular}
\end{center}
\end{table}

In the following, we mainly focus on the OZI-allowed two-body strong decay behaviors of these discussed kaons, by which we not only test these possible assignments to the observed kaons, but also provide more abundant predictions of higher radial and orbital excitations in the kaon family.

\subsection{$S$-wave kaons}

Since $K(498)$ and $K^*(892)$ were established to be the $1^1S_0$ and $1^3S_1$ states in the kaon family, respectively, in this work we do not discuss them, but present the phenomenological analysis of the $2S$ and $3S$ states.

\iffalse
For $K(498)$, its two-body OZI-allowed decays are forbidden. Thus, in this work we do not discuss this state.
\par
$K^*(892)$ is considered as $1^3S_1$ state and has only one decay mode $K^*(892)\rightarrow K\pi$. We can find that the widths given in both this work and Ref. \cite{Barnes:2002mu} are smaller than experimental data.
\fi

\subsubsection{$2S$ states}
%\paragraph{$K(1460)$ as $2^1S_0$ state }

As the candidate of the $2^1S_0$ state, the $K(1460)$ was listed in PDG. If further checking the experimental data, we find that the $K(1460)$ was only reported in Refs. \cite{Daum:1981hb,Brandenburg:1976pg}. However, in the past thirty years, further experiment of the $K(1460)$ was missing, which is the reason why the $K(1460)$ was removed from the its summary table of PDG.

In Table \ref{tab:S}, we give the information on the partial and total decay widths of the $K(1460)$ as an $2^1S_0$ state, in which one can find
the $K(1460)$ mainly decays into $K^*\pi$, $K\rho$, and $K\omega$. Here, our results are larger than the experimental data for $K^*\pi$, $K\rho$, and the total width. If establishing the $K(1460)$ to be an $2^1S_0$ state, we need to clarify these different between our calculation and experimental data.  We expect an independent experiment to confirm the observation of the $K(1460)$. Especially, we suggest precise measurement of the resonance parameters and partial decay widths of the $K(1460)$.

%\paragraph{Is $K^*(1410)$  $2^3S_1$ state? }

In PDG, the $K^*(1410)$ is possible candidate of the $2^3S_1$ state. However, we must face the following puzzling facts: (1)
the mass of the $K^*(1410)$ is smaller than that of the $K(1460)$. Usually, an $2^3S_1$ state has mass higher than that of an $2^1S_0$ state. In addition, we also notice the theoretical results of the mass of an $2^3S_1$ state, i.e.,
papers of Refs. \cite{Godfrey:1985xj,Ebert:2009ub} and this work give the mass of an $2^3S_1$ state to be 1579, 1675, 1548, and 1580 MeV, respectively, all of which larger than the experimental data, 1414 MeV, if $K^*(1410)$ is an $2^3S_1$  state. Thus, we need to understand why there exists such puzzling mass relation for the $K^*(1410)$ and $K(1460)$. (2) If the $K^*(1410)$ together with the $\rho(1450)$, $\omega(1420)$, and $\phi(1680)$  forms an $2^3S_1$ nonet, one can notice that the mass of the $K^*(1410)$ as an $2^3S_1$ state is a bit small which was also indicated in Ref. \cite{Barnes:2002mu}.

\begin{figure}[htbp]
\centering%
\begin{tabular}{c}
\scalebox{0.55}{\includegraphics{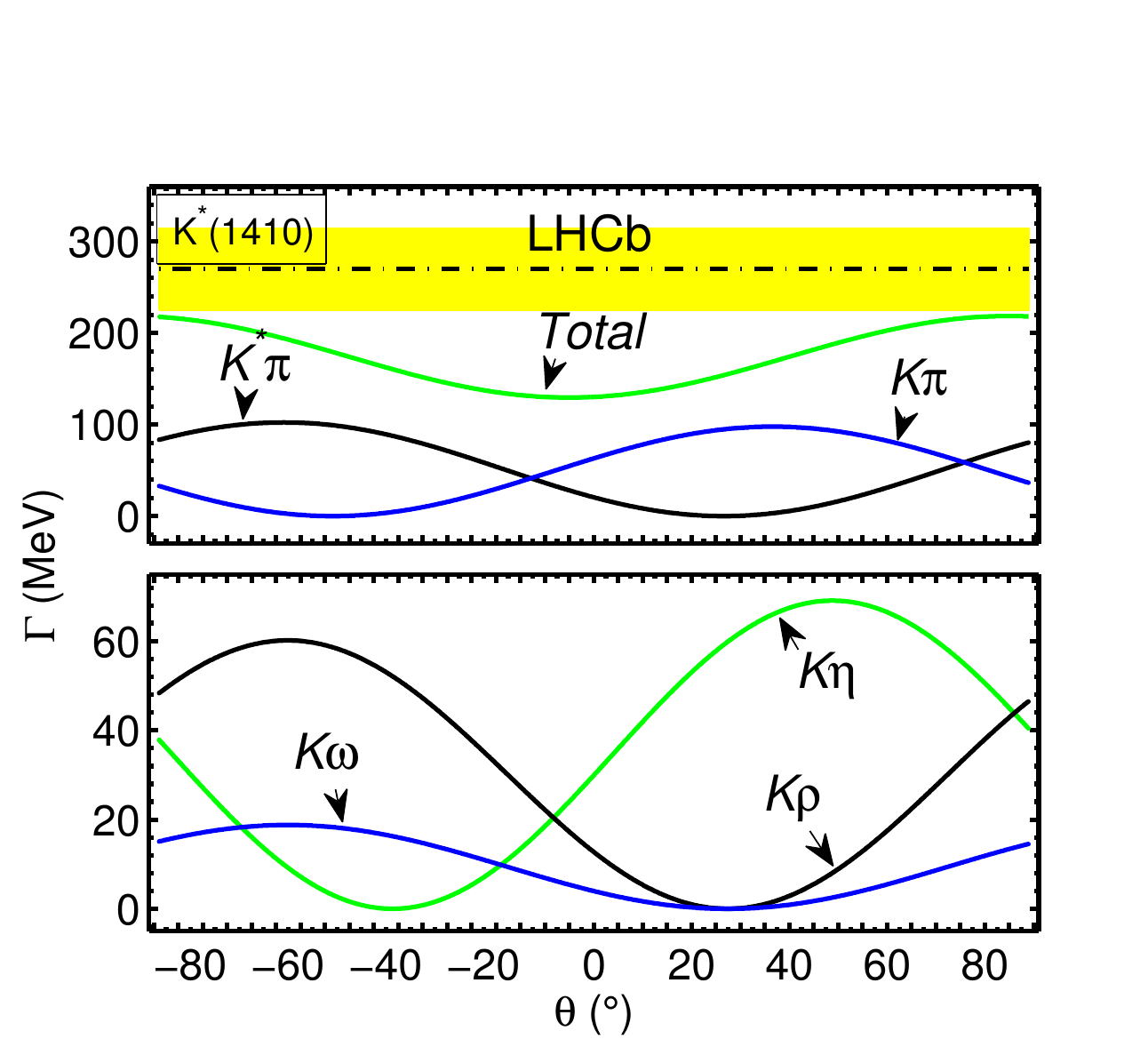}}
\end{tabular}
\caption{The $\theta_{sd}$ dependence of the total and partial decay widths of the $K^*(1410)$. Here, the dot-dashed line is the experimental value from LHCb \cite{Aaij:2015lsa}.}
 \label{fig:K1410}
\end{figure}

In Table \ref{tab:S}, the obtained partial and total decay widths of the $K^*(1410)$ as an $2^3S_1$ are given, where we also  compared our result with the experimental data.
The main decay modes of the $K^*(1410)$ include the $K\pi $, $K^*\pi$, $K\rho$, and $K\eta$ channels. The obtained total decay width of the $K^*(1410)$ is consistent with the experiment result.
{{We also notice that the ratio ${\Gamma_{K\pi}}/{\Gamma_{Total}}$ obtained in this work is a little bit larger than the experimental value $({\Gamma_{K\pi}}/{\Gamma_{Total}} =(6.6\pm1\pm0.8)\%)$.
The above result is gotten by assuming the $K^*(1410)$ as an $2^3S_1$ pure state. In fact, the $K^*(1410)$ could be a mixture of $2^3S_1$ and $1^3D_1$ states. Thus, in the following, we further
discuss such an S-D mixing effect on the ratio ${\Gamma_{K\pi}}/{\Gamma_{Total}}$ of the $K^*(1410)$.
The $K^*(1410)$ and $K^*(1680)$ as the mixture of $2^3S_1$ and $1^3D_1$ states can be expressed as
\begin{equation}
\left( \begin{array}{c}  |K^*(1410)\rangle \\ |K^*(1680)\rangle \end{array} \right) =
\left( \begin{array}{cc} \cos{\theta_{sd}} & \sin{\theta_{sd}} \\
                         -\sin{\theta_{sd}} & \cos{\theta_{sd}} \end{array} \right)
\left( \begin{array}{c} |1^3D_1\rangle\\ |2^3S_1\rangle \end{array} \right),
\label{sdmixing}
\end{equation}
where $\theta_{sd}$ denotes the mixing angle. Under this scenario, we present the decay behavior of
the $K^*(1410)$ dependent on $\theta_{sd}$ as shown in Fig. \ref{fig:K1410}.
The result shows that the experimental total width  \cite{Aaij:2015lsa} of the $K^*(1410)$ can be described when $\theta_{sd}$ is taken as $\sim 90^\circ$ or $\sim -90^\circ$, which supports the $K^*(1410)$ as a pure $2^3S_1$ state. We need to emphasize that the branching ratio ${\Gamma_{K\pi}}/{\Gamma_{Total}}$ becomes larger when $|\theta_{sd}|$ becomes smaller. Thus, the S-D mixing effect on $K^*(1410)$ state is not obvious if describing the experimental data. Of course, we must admit that there still exists a small difference between theoretical and experimental results for the ratio ${\Gamma_{K\pi}}/{\Gamma_{Total}}$. }}

Finally, we give a conclusion for the $K^*(1410)$.
The mass of $K^*(1410)$ as a $2^3S_1$ state is relatively small, and there exists some disagreement in the branching ratios with experiments.
Obviously, confirmation of this state assignment needs more experimental information and theoretical study.

 \begin{table}[htbp] \renewcommand{\arraystretch}{1.2}
\caption{The strong decay widths of the $2S$ and $3S$ states. The unit of the width is MeV. \label{tab:S}}
\begin{center}
\begin{tabular}{cccccccccccc}
 \toprule[1pt] \toprule[1pt]
 {States} &Channels&This work&Experiment \\
 \midrule[1pt]
% $K^*(892)(1^3 {S}_1 )$& $K\pi$&17.9 &21&$\sim$48~\cite{Agashe:2014kda}& \\

   $K(1460)\,[2^1 {S}_0)]$
                       &$K^*\pi $&248&$\sim 109$~\cite{Daum:1981hb}\\
                      & $K\rho $& 161&$\sim 34$~\cite{Daum:1981hb}\\
                      &$K\omega$ &51.2 &--\\
                      &$K^*\eta$ &8.0 &--\\
                     &Total& 468&$\sim 260$~\cite{Daum:1981hb}\\
                                                           \midrule[1pt]
   $K^*(1410)\,[2^3 {S}_1]$

                                  &$K^*\pi $&81.8 &--\\
                                  & $K\rho$ &47.4&-- \\
                                   &$K\omega$ &14.8 &--\\
                                   &$K\pi$ &34.7&--\\
                                   &$K\eta$ &35.4 &--\\
                              & Total& 214&$232\pm21$~\cite{Agashe:2014kda} \\
                &$\Gamma_{K\pi}/\Gamma_{Total}$&16.2 \%&$(6.6\pm1\pm0.8)\%$\\
\midrule[1pt]

  $K(1830)\,[3^1 {S}_0]$

                                 &$K^*(1410)\pi$& 105 &--\\
                                  &$ K^*\pi$ & 34.7&--\\
                                  &$K_0^*(1430)\pi$ &29.8&--\\
                                   &$K\rho $& 22.4&-- \\
                                  &$K_2^*(1430)\pi$& 21.7&--\\

                          &$ K^*\rho$ & 17.4&--\\

                                  &$K\omega$&7.07&--\\
                                   &$K^*\omega$   & 6.0 &--\\
                                   &$K^*\eta$   &1.26&--\\

                                    &$K\phi$   &0.018&--\\

                                    &Total&245 &$\sim250$~\cite{Agashe:2014kda} \\\midrule[1pt]

                                   $3^3S_1$&$K^*{(1410)}\pi  $ & 62.3 &--\\
 & $K^* \pi  $& 44.2 &--\\&
$ {K} \rho $ & 36.8 &--\\&
 $K_2^* \pi $ & 29.4 &--\\&
 ${K} \pi $ & 23.3 &--\\&
 $  K(1460)\pi$ & 20.8 &--\\&
 ${K} {\eta (1295)}$ & 18.5 &--\\&
 ${K} a_2 $& 15.8 &--\\&
 ${K} {\pi (1300)}$ & 14 &--\\&
 ${K} \eta $ & 13.1 &--\\&
 ${K} \omega$  & 12 &--\\&
 $  K_1(1650)\pi $& 11.8 &--\\&
 ${K} b_1$ & 11.6 &--\\&
 $\eta K_1{(1400)}$  & 7.93 &--\\&
 $K_1 \eta $ & 7.93 &--\\&
 ${K} a_1$ & 7.76 &--\\&
 ${K} f_2$ & 7.65 &--\\&
 ${K} {\omega (1420)}$ & 7.62 &--\\&
 ${K} f_1{(1420)}$ & 7.18 &--\\&
 $K^* \phi $ & 5.54 &--\\&
 ${K} {\rho (1450)} $& 5.46 &--\\&
 ${K} h_1$ & 4.22 &--\\&
$  K_1{(1400)\pi} $ & 3.82 &--\\&
$ K_1 \pi  $& 3.82 &--\\&
$ K^* \rho $ & 3.38 &--\\&
$ {K} f_1 $& 2.92 &--\\&
 ${K} \phi$  & 2 &--\\
 &
 Total& 393 &--\\
\bottomrule[1pt]\bottomrule[1pt]
\end{tabular}
\end{center}
\end{table}
\subsubsection{$3S$ states}

%\paragraph{$K(1830)$ as $3^1S_0$ state }

Although the $K(1830)$ is not listed in the summary table of PDG, we
still select the $K(1830)$ as a possible candidate of the $3^1S_0$ state and study its decay behavior.

In Table \ref{tab:S}, the partial and total decay widths of the $K(1830)$ as a $3^1S_0$ state are shown.
Our results show that the largest decay width of $K(1830)$ is given by the channel $K^*(1410)\pi$ instead of $K^*\rho $ given by  Ref. \cite{Barnes:2002mu}.
The other main decay channels contain $K^*\pi$, $K_0^*(1430)\pi$, $K\rho$, $K_2^*(1430)\pi$, and $K^*\rho$. The total width of the theory agrees with the experimental data. Our prediction of the decay information on this state will be helpful for the future experimental  study, since there exists {only two experimental studies on $K(1830)$ until now}.

At present, the $3^3S_1$ state in the kaon family is still absent. Thus, in this work we predict its decay property, where we take the predicted mass of
 the $3^3S_1$ state by the {MGI} model as the input.
 The results shown in Table \ref{tab:S} indicate that its important decay modes are $\pi  K^*(1410)$, $K^* \pi $,
$K\rho$, and  $K_2^* \pi$, $K \pi $,
 $\pi  K(1460)$, ${K} {\eta (1295)}$. Additionally, $K a_2 $ also have sizable contribution to the total width.
This predicted decay information is useful to the future experimental search for this missing state.

\subsection{$P$-wave kaons}
%%%%%%%%%%%%%%%%%%%%%%%%%%%%%%%%%%%%%%%%%%%%%%%%%%%%%%%%%%%%%%%%%%%%%%%%%%%%%%%%%%%%%%%%%%%%%%%%%%%%%%%%%%%%%%%%%%%%%%%%%%%%
%\paragraph{$K_0^*(1430)$ as $1^3P_0$ state}

\subsubsection{$1P$ states}

In Table \ref{tab:P}, we show the allowed decay channels of the $K_0^*(1430)$, and the corresponding partial and total decay widths. Here, its dominant decay channel of the $K_0^*(1430)$ is $K\pi$, which has decay width $314$ MeV, which is comparable with the experiment data $(267\pm36)$ MeV listed in PDG \cite{Agashe:2014kda}. % and better than that of Ref.~\cite{Barnes:2002mu}.
Besides, the $K\eta$ decay channel also has sizable
contribution to the total decay width of $K_0^*(1430)$. In addition, the obtained total decay width is consistent with the experimental measurement just shown in Table \ref{tab:P}. The above study indicates that the $1^3P_0$ assignment to the $K_0^*(1430)$ is suitable.

%%%%%%%%%%%%%%%%%%%%%%%%%%%%%%%%%%%%%%%%%%%%%%%%%%%%%%%%%%%%%%%%%%%%%%%%%%%%%%%%%%%%%%%%%%%%%%%%%%%%%%%%%%%%%%%%%%%%%%%%%%%%
%\paragraph{$K_2^*(1430)$  as $1^3P_2$ state }
\par
The $K_2^*(1430)$ together with $a_2(1320)$, $f_2(1270)$, and $f_2^\prime(1525)$ may form a $1^3P_2$ nonet.
In Table \ref{tab:P}, we give the partial decay widths of the $K_2^*(1430)$.
It dominantly decays into $K\pi$ and $K^*\pi$,
while the $K\rho$, $K\omega$, and $K\eta$ modes also have sizable contributions in which $K\eta$ was already observed in experiment \cite{Agashe:2014kda}.
According to Table \ref{tab:P}, we can find that our results are consistent with experimental data. Thus, the $K_2^*(1430)$ as a $1^3P_2$ state in the kaon family can be supported by our study of its decays.

 \begin{table}[htbp]
 \renewcommand{\arraystretch}{1.2}
\caption{The decay widths of three $P$-wave states. The unit of the width is MeV. \label{tab:P}}
\begin{center}
\begin{tabular}{cccccccccccc}
\toprule[1pt]\toprule[1pt]
 {States} &Channels&This work&Experiment \\
\midrule[1pt]

$K_0^*(1430)\, [1^3 {P}_0]$& $K\pi$ & 314&$267\pm36$~\cite{Aston:1987ir} \\
 & K$\eta$&2.87&-- \\
  &Total& 318&$270\pm80$~\cite{Agashe:2014kda}\\
\midrule[1pt]
 $K_2^*(1430) \,[1^3 {P}_2]$&$K\pi $ &51.5&$48.9\pm1.7$~\cite{Agashe:2014kda,Blundell:1996as}\\
                      &$K^*\pi$  &20.4&$24.8\pm1.7$~\cite{Agashe:2014kda,Blundell:1996as}\\
                      &$K\rho$&6.13&$8.7\pm0.8$~\cite{Agashe:2014kda,Blundell:1996as}\\
                      &$K\omega$&1.82 &$2.9\pm0.8$~\cite{Agashe:2014kda,,Blundell:1996as}\\
                      &$K\eta $ &0.0665&$0.15^{+0.37}_{-0.1}$~\cite{Agashe:2014kda}\\
                                      & Total& 80.1&$98.5\pm2.9$~\cite{Agashe:2014kda} \\
                                       \midrule[1pt]
$K_0^*(1950) \,[2^3 {P}_0]$&
                        $K\pi$ &105 &--\\
                        &$ K^*\rho $& 254&-- \\
                             &$K\pi(1300)$ &190 &--\\
                              &$K(1460)\pi$&121 &--\\
                              &$K a_1$&69.1 &--\\
                               &$K b_1$&64.9 &--\\
                                 &$K_1(1270) \pi$&183&--\\

                                 &$K_1(1270) \eta$&6.58&--\\

                               &$K_1(1400) \pi$&5.98 &--\\

                                &Total&1000&$201\pm34\pm79$~\cite{Aston:1987ir}\\
 &$\Gamma_{K\pi}/\Gamma_{Total}$&  10.5 \%&$(52\pm8\pm12)\%$~\cite{Aston:1987ir}\\
                                                         \bottomrule[1pt]\bottomrule[1pt]

%$ K(1630)(3^1 {S}_0) $
                                     %&$K \rho$ &0.485&-- \\
                                     %  &$K^*\pi$&3.09 &--\\
                                       % &$K\omega$&0.119 &--\\
                                        %&$ K^* \eta $ & 0.464&-- \\
                                        %&$K^*(1410)\pi$&11.7 &--\\
                                       % &$K_0^*(1430)\pi$ &14.3 &--\\
                                % &Total & 30.1&--&$16^{+19}_{-16}$ \\

%\hline
%\hline
\end{tabular}
\end{center}
\end{table}

The $K_1(1270)$ and $K_1(1400)$ as the $1P$ and $1P^\prime$  states respectively satisfy
\begin{equation}
\left( \begin{array}{c}  |K_1(1270)\rangle \\ |K_1(1400)\rangle \end{array} \right) \approx
\left( \begin{array}{cc} \cos{\theta_{1P}} & \sin{\theta_{1P}} \\
                         -\sin{\theta_{1P}} & \cos{\theta_{1P}} \end{array} \right)
\left( \begin{array}{c} |1^1P_1\rangle\\ |1^3P_1\rangle \end{array} \right),
\label{kmixing}
\end{equation}
where $\theta_{1P}$ denotes the mixing angle, which makes us discuss the $\theta_{1P}$ dependence of the partial and total decay widths of the $K_1(1270)$ and $K_1(1400)$.

\begin{figure}[htbp]
\begin{center}
\begin{tabular}{c}
\scalebox{0.6}{\includegraphics{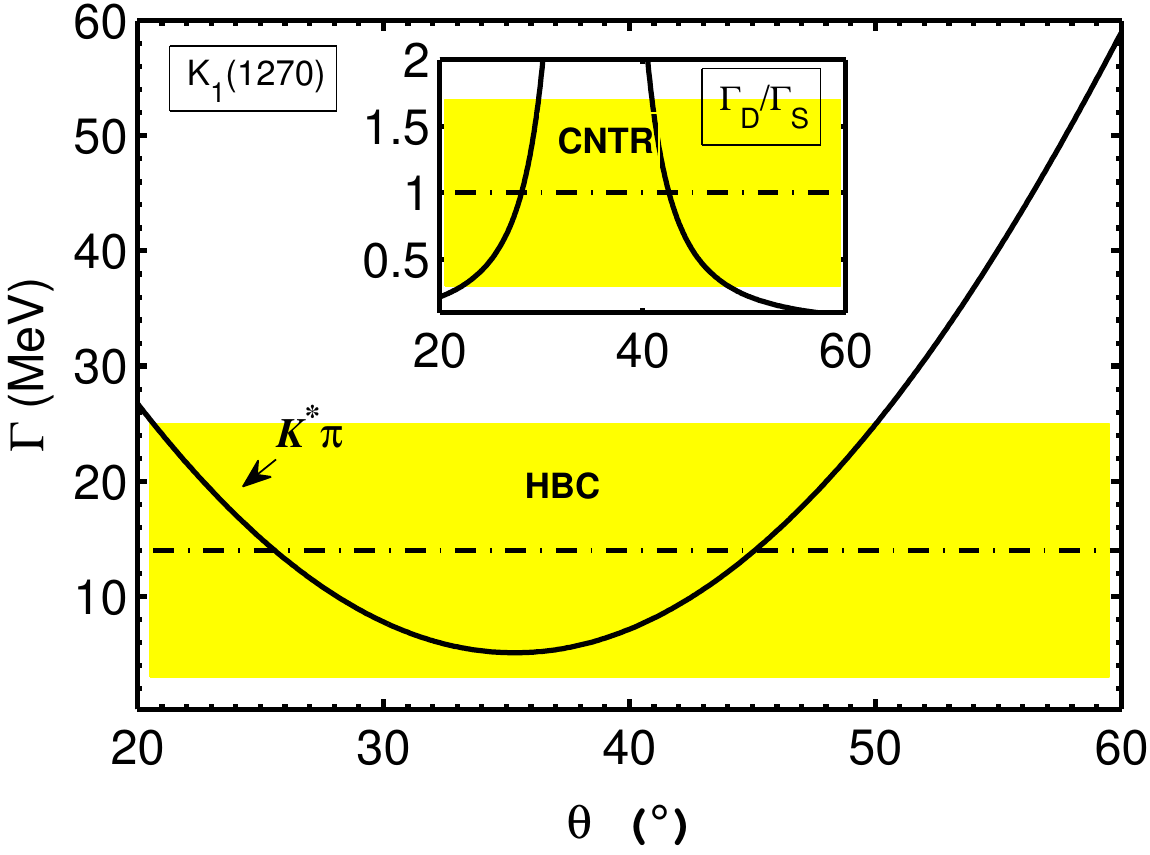}}
\end{tabular}
\caption{$\theta_{1P}$ dependence of the $K_1(1270)\to K^*\pi$ decay width.}
 \label{K11270}
 \end{center}
\end{figure}

According to Fig.~\ref{K11270} which { describes mixing angle $\theta_{1P}$ dependence of the $K_1(1270)$ decay width}, we find that $\theta_{1P}$ should be taken as either $22.5^\circ\sim 29^\circ$~ or $41.5^\circ\sim 48^\circ$ by fitting the CNTR data of $\Gamma_{K^*\pi}$ \cite{Daum:1981hb}, which is fortunately in the same range when fitted with the ratio ${\Gamma(K^*\pi)}_{D-wave}/{\Gamma(K^*\pi)_{S-wave}}=1\pm0.7$ ~\cite{Daum:1981hb}. Here, the central value of this mixing angle is
$\theta_{1P} \approx 25^\circ$ or~$45^\circ$.

\begin{figure}[htbp]
\begin{center}
\begin{tabular}{c}
\centering%
\scalebox{0.6}{\includegraphics{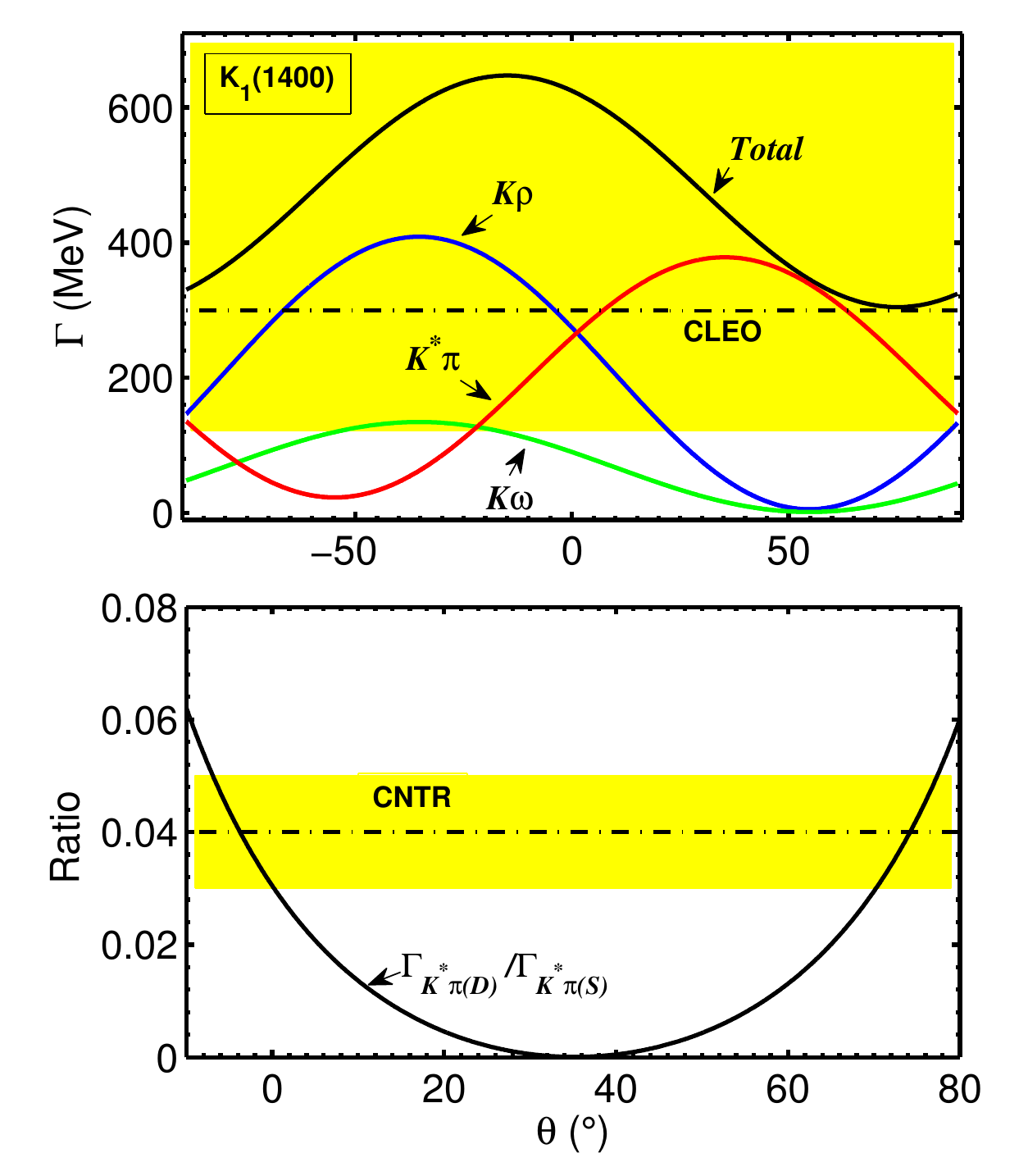}}
\end{tabular}
\caption{$\theta_{1P}$ dependence of the partial, the total decay widths and ratio $\Gamma_{K^*\pi(D)}/\Gamma_{K^*\pi(S)}$ of the $K_1(1400)$ .}
 \label{K11400}
 \end{center}
\end{figure}

We further investigate the decays of the $K_1(1400)$. CNTR~\cite{Daum:1981hb} also gave ${\Gamma_{K^*\pi}}/\Gamma_{Total}=(94\pm6)\%$ \cite{Carnegie:1976cs} for the $K_1(1400)$, by which we obtain $38^\circ<\theta_{1P} < 68^\circ$ with central value
  $\theta_{1P}=45^\circ$, where the details can be found in Fig. \ref{K11400}\footnote{
CNTR~\cite{Daum:1981hb} gave the $\Gamma_{K^*\pi(D)}/\Gamma_{K^*\pi(S)}=0.04\pm0.01$ as well, which gives the angle range
  $-7^\circ\sim 0^\circ$ or $70^\circ\sim 78^\circ$ if we fit with the ratio, which conflicts with the previous discussion.}
Hence, the above analysis shows that the mixing angle $\theta_{1P}$ favors $45^\circ$ which agrees with the conclusion made in Refs. \cite{Barnes:2002mu,Blundell:1996as} but disagrees with Refs.~\cite{Cheng:2003sm,Li:2006we,Tayduganov:2011ui}, in which they obtained~$\theta_{1P}=34^\circ$ and $\sim60^\circ$, respectively.
    \par

\subsubsection{$2P$ states}
%\paragraph{$K_0^*(1950)$ as $2^3P_0$ state}

As shown in Table \ref{tab:kmassground1},
papers of Refs. \cite{Godfrey:1985xj,Ebert:2009ub,Barnes:2002mu} and this work
give the mass of the $2^3P_0$ state 1.890, 1.791, 1.850, and 1.829 GeV, respectively, which are all smaller than the experimental value 1945 MeV if the $K_0^*(1950)$ is assumed to be a $2^3P_0$ state.
Under the assignment of the $2^3P_0$ state to the $K_0^*(1950)$, we study the strong decay behavior of the $K_0^*(1950)$, which is presented in
Table \ref{tab:P}.

Our results show that the $K^*\rho$ mode is its dominant decay channel.
Its total decay width can reach up to 1000 MeV which is 5 times larger than the experimental value 200 MeV.
We also notice the result of the total decay width of a $2^3P_0$ state given by Ref. \cite{Barnes:2002mu}, which is two times larger than the experimental value, where they use a smaller phase space (their mass of a $2^3P_0$ state is 1850 MeV).
We also obtain the branching ratio ${\Gamma_{K\pi}}/{\Gamma_{Total}}=6.4\%$, which is close to $10.5\%$ calculated by Ref. \cite{Barnes:2002mu}, but are smaller than the experimental value $52\%$.
Besides, we also confirm that $K_1(1270)\pi$ has sizable contribution to the width of the $K_0^*(1950)$ \cite{Barnes:2002mu}.
It is obvious that there exists a difference between the present theoretical and experimental results.
Until now, the $K_0^*(1950)$ has not been established in experiment since this state was omitted from the summary table of PDG \cite{Agashe:2014kda}. For clarifying it, we suggest further experimental study of the $K_0^*(1950)$, where its resonance parameter and partial decay widths are crucial information.

Then, we discuss possibility of two different assignments to the $K_2^*(1980)$ from two aspects, mass and decay information.
%\subparagraph{Mass analysis}
%\subparagraph{As $1^3F_2$ state}
In 1987, LASS reported a structure in the reaction $K^- p \to \bar{K}^0 \pi^+ \pi^- n$%, which probably includes the kaons with $1^3F_2$ or $2^3P_2$ quantum numbers in the final state
~\cite{Aston:1986jb}, and the obtained resonance parameter are $M=(1973\pm8\pm25)$ MeV and $\Gamma=(373\pm33\pm60)$ MeV. This is the particle called $K_2^*(1980)$ listed in PDG \cite{Agashe:2014kda}.
Barnes {\it et al.}~\cite{Barnes:2002mu} have the viewpoint that the $K_2^*(1980)$ is a $1^3F_2$ state, and give the total width 300 MeV. However,
our results show that the mass of a $1^3F_2$ state is about 2093 MeV. Thus, the mass of the $K_2^*(1980)$ is a bit small if $K_2^*(1980)$ is a $1^3F_2$ state, which can be supported by another fact, i.e., as an iso-vector $1^3F_2$ state, $a_2(2030)$ is well established in Ref. \cite{Pang:2014laa}. {In the same $1^3F_2$ nonet, the meson which contains one $s$ quark is heavier than the mesons which only contain $u/d$ quarks.}
{Along this line,} the mass of the $1^3F_2$ state in the kaon family should be heavier than 2030 MeV.

Assuming the $1^3F_2$ state assignment to $K_2^*(1980)$, we illustrate its decay behavior.
The present work (see Table \ref{tab:1F2}) shows that the $K_1(1270)\pi$ is the dominant decay channel when we treat the $K_2^*(1980)$ as a $1^3F_2$ state, even though the channel is not observed in experiments.
$K_2(1770)\pi$, $Kb_1$, $Ka_1$, $K\pi$, $K\rho$, and $K^*\pi$ modes, among which $K\rho$ and $K^*\pi$ have been reported in the experiment \cite{Agashe:2014kda}, also have sizable contributions, where we take $\theta_{1D}=-39^\circ$.  Our prediction for the channels $K_1(1270)\pi$, $K_2(1770)\pi$, $Kb_1$, $Ka_1$, and $K\pi$ will be helpful for the
experimental test of this assignment.

Besides the assignment of the $1^3F_2$ state to the $K_2^*(1980)$, there exists another possibility, the $K_2^*(1980)$ as a $2^3P_2$ state. The analysis of mass spectra in Refs. \cite{Godfrey:1985xj,Ebert:2009ub,Barnes:2002mu} and this work
shows that  the mass of a $2^3P_2$ state is 1938, 1896, 1850, and 1870 MeV, respectively.
Thus, the experimental mass value of $K_2^*(1980)$ is a bit larger as a $2^3P_2$ state. If the $K_2^*(1980)$ is a $2^3P_2$ state, its main decay modes are $K^*\rho$, $K\pi$, $K^*\pi$, $K\rho$, $K^*\eta$, $K\eta^\prime$, and $K^*\omega$.
Besides the $K\rho$ and $K^*\pi$ modes, one notices that $Kf_2$ has been observed  in experiments which has a sizable contribution in theory.
Hence, the $K_2^*(1980)$ as a $2^3P_2$ state is also a possible assignment.

Just presented above, we discuss two assignments to the $K_2^*(1980)$, where the decay behaviors of the $K_2^*(1980)$ under two assignments are different. Thus,
we should combine further experimental decay information of the $K_2^*(1980)$ with our results to determine which possibility of its assignments we should take.
%Finally, let us draw a rough conclusion for the $K_2^*(1980)$.
%According to the mass analysis, the mass of the $K_2^*(1980)$ is a bit small when assigned to a $1^3F_2$ state and a bit large when assigned to a $2^3P_2$ state.
%According to the decay information, the $K_2^*(1980)$ is in favor of a $2^3P_2$ state. We still, however, need more experimental information to test our assignment for $K_2^*(1980)$.
%What is more important is that  we give the prediction that the partial widths of
%$K_2^*(1425)\pi$, $K^*(1410)\pi$,
%$K^*\omega$, $K^*\rho$, and $K^*\eta$  treating the $K_2^*(1980)$ as a $1^3F_2$ state will be much larger than  those of the case of a $2^3P_2$ state.
%The experimental study of these decay modes will be helpful to test this two assignments to the $K_2^*(1980)$.

 \begin{table*}[htbp]
 \renewcommand{\arraystretch}{1.2}
\caption{The strong decay widths of $K_2^*(1980)$, where the values in brackets and without brackets in the third and fourth columns represent those for the $K_2^*(1980)$ as the $1^3 {F}_2$ and $2^3P_2$ states, respectively.
The unit of the width is MeV. \label{tab:1F2}}
\begin{center}
 %\resizebox{!}{5.5cm}{
\begin{tabular}{cccccccccccc}
\hline\hline
 States&Channel&This work&Ref.~\cite{Barnes:2002mu}&Experiment\\
 \hline
 $K_2^*(1980)$
  & $K_1(1270) \pi$  &36.5(55.7) &6(79)& -- \\
    & $K_2(1770)\pi$  &0.473(5.67)&--(61)\\
    & $K b_1$   &24.1(36.7) &8(50)& -- \\
    & $K a_1$  &11.8(10.8)&3(26)& -- \\

   & $K_1(1270) \eta$   &6.31(5.57) &1(22)& -- \\
    & $K \pi$   & 0.788(17.5) &44(20)& -- \\
     & $K h_1$  &11.8(9.00) &4(18)& --\\

     & $K \rho$   &10.6(18.3) &44(13)& -- \\
  & $K^* \pi $ &10.2(16.2)&47(13) & -- \\
  & $K \eta ' $ &3.50(6.51) &15(15)& -- \\

  & $K^* \eta  $ & 24.6(6.80) &26(11)& -- \\
   & $K_2^*(1425) \pi$   &61.3(17.4) &15(8)& -- \\

   & $K^* \rho $  &42.1(17.1) &78(8)& -- \\

   & $K f_1(1282)$  &2.93(2.67) &1(7)&-- \\
    & $K a_2$  &28.4(11.8)&3(7) & -- \\
    & $K f_1(1426)$  &0.454(0.32) &--(6)&-- \\
    & $K \phi$   &11.1(3.31) &12(6)& -- \\

  & $K \omega $  & 3.60(6.09)&14(4)& -- \\

  & $K f_2(1270)$  &12.1(5.55)&3(3) & -- \\

& $K^* \omega $  &22.0(5.42)&27(3) & -- \\
       & $K^* \phi$   &41.0(0.109) &--(1)& -- \\
    & $K_1(1400) \pi$&7.56(7.32) &11(0)&--  \\

  & $K^*{(1410)} \pi$  &  63.9 (2.03)&5(0)& -- \\
  & $\pi  K(1460)$ &  27.9(4.31)& 2(0)& -- \\
   & $K\pi(1300)$  &15.0(5.57)&0(0)&-- \\
   &Total & 480(278)&370(283)&$373\pm33\pm60$~\cite{Aston:1986jb} ($180\pm 70$ \cite{Tikhomirov:2003gg}) \\
 &$\Gamma_{K\rho}/\Gamma_{K^*\pi}$ &1.04(1.14) &0.94(1)&$1.49\pm0.24\pm0.09$~\cite{Aston:1987ir} (--)\\
\hline
\hline
\end{tabular}
\end{center}
\end{table*}

The $K_1(1650)$ and its partner $K_1(2P^\prime)$ satisfy
\begin{equation}
\left( \begin{array}{c}  |K_1(1650)\rangle \\ |K_1(2030)\rangle \end{array} \right) \approx
\left( \begin{array}{cc} \cos{\theta_{2P}} & \sin{\theta_{2P}} \\
                         -\sin{\theta_{2P}} & \cos{\theta_{2P}} \end{array} \right)
\left( \begin{array}{c} |2^1P_1\rangle\\ |2^3P_1\rangle \end{array} \right).
\label{kmixing}
\end{equation}
%where $K_1(2030)$  is obtained as follows.
\begin{figure}[htbp]
\centering%
\begin{tabular}{c}
\scalebox{0.60}{\includegraphics{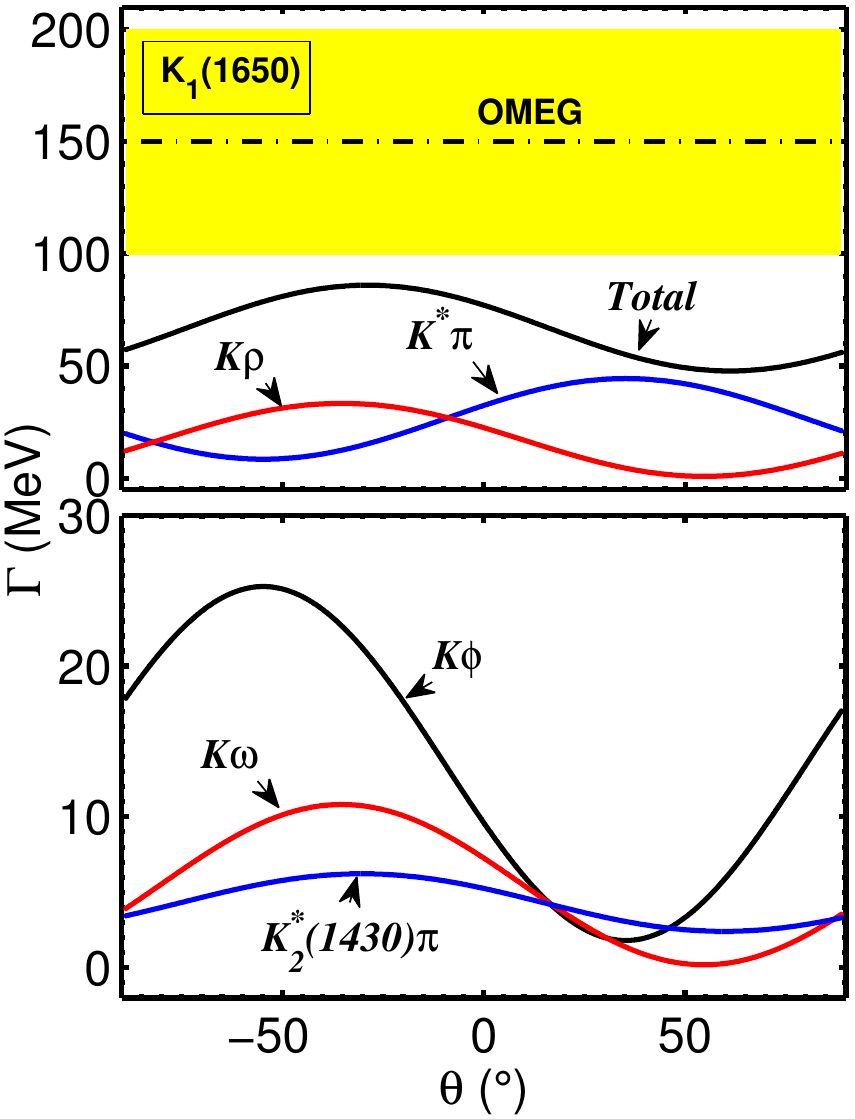}}
\end{tabular}
\caption{$\theta_{2P}$ dependence of the decay widths of the $K_1(1650)$ as a $2P$ state, where the dot-dashed line is the experimental value of OMEG \cite{Frame:1985ka}.}
 \label{K11650}
\end{figure}
\begin{figure}[htbp]
\centering%
\begin{tabular}{c}
\scalebox{0.50}{\includegraphics{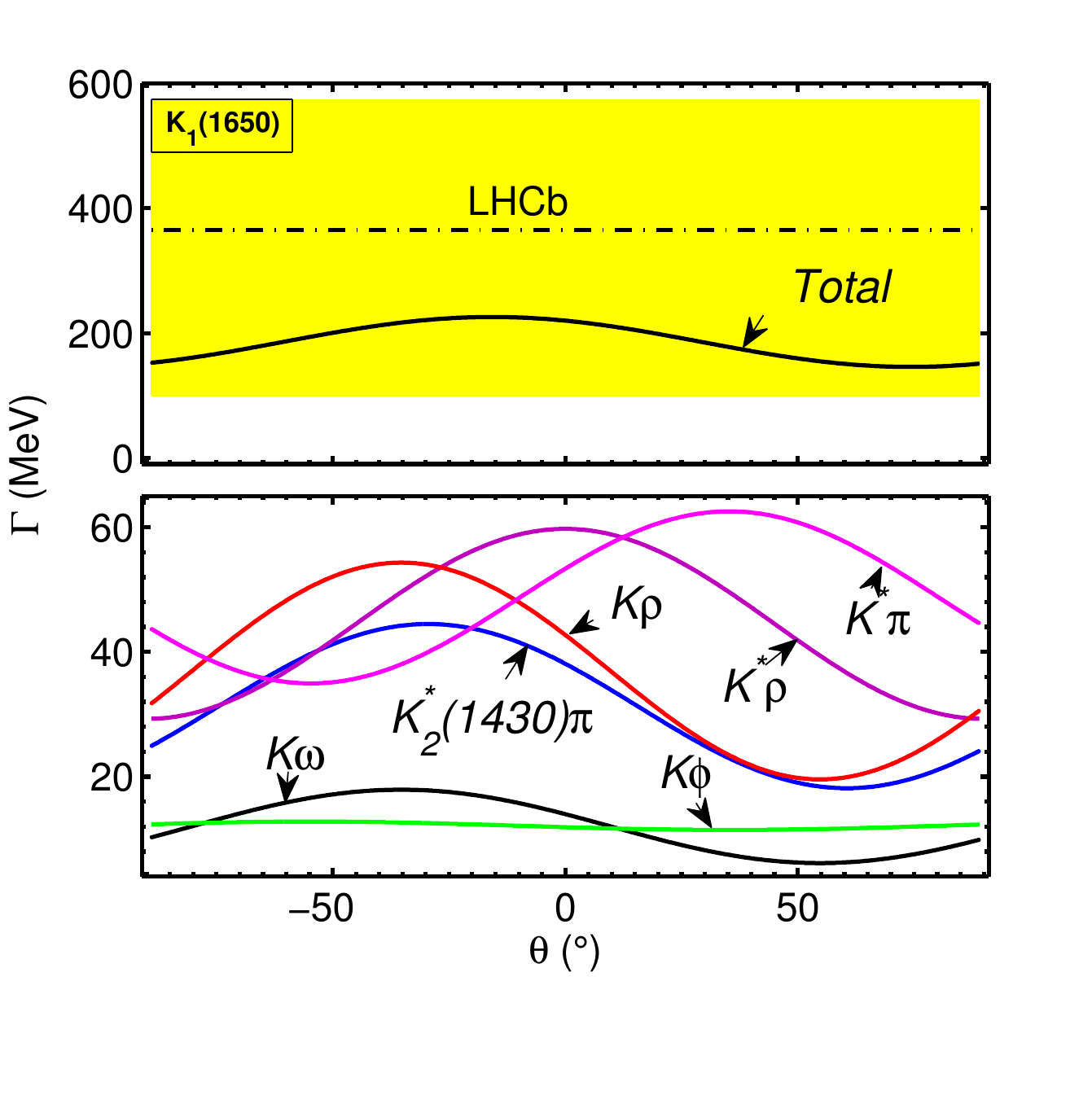}}
\end{tabular}
\caption{$\theta_{2P}$ dependence of the decay widths of the $K_1(1650)$ as a $2P$ state with the measured center mass of 1793 MeV from LHCb \cite{Aaij:2016iza}, where the dot-dashed line is the experimental value of LHCb \cite{Aaij:2016iza}.}
 \label{K11790}
\end{figure}

In Fig. \ref{K11650}, we show the partial and total decay widths of the $K_1(1650)$ depending on the mixing angle $\theta_{2P}$ if the mass of the $K_1(1650)$ is adored to be $M=1650\pm50$ MeV \cite{Agashe:2014kda}.
{{Since the decays of $K_1(1650)$ into $K\pi\pi$ and $K\phi$ were observed in experiments \cite{Daum:1981hb,Frame:1985ka,Armstrong:1982tw,Aaij:2016iza}, we can roughly conclude that  $\theta_{2P}$ is probably less than zero as seen from Fig. \ref{K11650}, where $K\rho$, $K\phi$ and $K^*\pi$ have sizable contributions to the total decay width of the $K_1(1650)$ in our calculation.

In experiment, the $K_1(1650)$ is also not well established since this state is omitted from the summary table of PDG \cite{Agashe:2014kda}. More experimental and theoretical efforts are necessary to establish the $K_1(1650)$.
We notice new experimental information of the $K_1(1650)$ from LHCb \cite{Aaij:2016iza}, where the measured mass of the $K_1(1650)$ is {$(1793\pm59^{+153}_{-101})$} MeV which is about 150 MeV larger than experimental data given by Ref. \cite{Agashe:2014kda}. Taking the LHCb mass result as an input, we investigate the strong decay behaviors of the $K_1(1650)$ again, which are shown in Fig. \ref{K11790}. Here, $K^*\pi$, $K^* \rho$, $K\rho$ and $K_2^*(1430)\pi$ are dominant decay channels.
However, we cannot give further constraint on the mixing angle $\theta_{2P}$ by comparing an experimental width with our theoretical result due to a large experimental error of the LHCb experimental data.

In the following, we discuss the partner of the $K_1(1650)$.
Ref. \cite{Cheng:2011pb} gives the following equation about the mass relation between the pure states and physical states
%\begin{equation}
\begin{eqnarray}\label{mpmpp}
m_{K_1(2^3P_1)}^2 = \cos({\theta_{2P}})^2 m_{K_1(2P^\prime)}^2+ \sin({\theta_{2P}})^2 m_{K_1(2P)}^2,\\\nonumber
m_{K_1(2^1P_1)}^2 = \sin({\theta_{2P}})^2 m_{K_1(2P^\prime)}^2+ \cos({\theta_{2P}})^2 m_{K_1(2P)}^2.
%\end{equation}
\end{eqnarray}
Substituting $m_{K_1(2^1P_1)}=1840$ MeV and $m_{K_1(2^3P_1)}=1861$ MeV given in Table \ref{tab:kmassground1} into Eq. (\ref{mpmpp}),
we obtain the  mass of the $K_1(1650)$ partner, {$m_{K_1(2P^\prime)}$}, about 2030 MeV and the $\theta_{2P} \approx \pm43^\circ$ if taking the mass of the $K_1(1650)$ as {$M=(1650\pm50)$} MeV \cite{Agashe:2014kda}.
Based on the above analysis of the $K_1(1650)$, we suggest $\theta_{2P} \approx -43^\circ$.
If taking the LHCb mass measurement \cite{Aaij:2016iza} of the $K_1(1650)$ as an input, the mass of the $K_1(1650)$ partner is estimated to be 1906 MeV. In Ref. \cite{Aaij:2016iza}, the resonance parameters of the $K_1(1650)$ partner are given, i.e., $M=(1968\pm65^{+70}_{-172})$ MeV and $\Gamma=(396\pm170^{+174}_{-178})$ MeV.
Considering the present status, we select the experimental mass (1968 MeV  \cite{Aaij:2016iza}) for the partner of the $K_1(1650)$ when discussing the decay behavior of the $K_1(2P^\prime)$ state just shown in Table \ref{tab:k12030}. Here, the calculated width of  the $K_1(2P^\prime)$  is about $(440-570)$ MeV,  which is comparable to the experimental data \cite{Aaij:2016iza}.
Its main decay modes are $K^*\rho$, $K_2^*\pi$, $K a_2$, $K^*\pi $, ${K} \rho $, $K^* \omega$, and $K f_2$. }}

 \begin{table}[htbp]
 \renewcommand{\arraystretch}{1.2}
\caption{The main strong decay widths of the $K_1(2P^\prime)$ state which as the partner of the $K_1(1650)$. Here, the, the mass of  the $K_1(2P^\prime)$ is taken as $1968$ MeV \cite{Aaij:2016iza}. $c=\cos\theta_{2P}$ and $s=\sin\theta_{2P}$.
The unit of the width is MeV. \label{tab:k12030}}
\begin{center}
 %\resizebox{!}{5.5cm}{
\[\begin{array}{cccccccccccc}
\hline\hline
 \text{Decay channel}&\text{Width}\\
 \hline
 \text{Total} &440 c^2+70.2 c s+385 s^2\\
 K^* \rho     & 166 c^2+111 s^2\\
 K_2^* \pi   &   71.5 c^2+57 c s+111 s^2\\
 {K} a_2    &59c^2+48. c s+80.8 s \\
 K^* \pi   &75.4 c^2+8.84 c s+72.3 s^2\\
 K \rho    &63.1 c^2+2.81 c s+64.1 s^2\\
 K^* \omega  &53.8 c^2+36 s^2 \\
 {K} f_2   &  22.4 c^2+10.5 c s+20.8 s^2\\
\hline
\hline
\end{array}\]
\end{center}
\end{table}

\subsection{$D$-wave kaons}
\subsubsection{1$D$ states}
%\paragraph{$K^*(1680)$ as $1^3D_1$ }
$K^*(1680)$ together with $\rho(1700)$ and $\omega(1650)$ forms a $1^3D_1$ nonet. Barnes {\it et al.}~\cite{Barnes:2002mu} predicted that this state should have the mass 1850 MeV, but we obtain 1.766 GeV which is closer to the experimental value 1.717 GeV.
The mass spectrum analysis supports $K^*(1680)$ as a $1^3D_1$ state.

{\iffalse
{{
In the other hand, when we consider the mixing of Eq. (\ref{sdmixing}), the analysis of $K^*(1410)$  supports  $K^*(1680)$ as a pure $1^3D_1$ state.
In Table \ref{tab:k1680}, the expressions of partial widths of $K^*(1680)$ with $\theta_{sd}$ dependence are given and experiments \cite{Agashe:2014kda} give two useful ratios of partial width of
$K^*(1680)$, $\frac{K\pi}{K^*\pi}=1.3^{+ 0.23}_{- 0.14}\%$,  $\frac{K\rho}{K\pi}=0.81^{+ 0.14}_{- 0.09}\%$. Comparing this information with Table \ref{tab:k1680}, we can conclude that the  mixing angle $\theta_{sd}$ is about $86^\circ-90^\circ$ which is consistent with the previous analysis of $K^*(1410)$.
}}

\begin{table}
\caption{The decay widths of the $K^*(1680)$ when considering S-D mixing. Here, $c=\cos\theta_{sd}$ and $s=\sin\theta_{sd}$. The unit of the width is MeV.}
\[\begin{array}{cc}
\hline\hline
Decay~channel&Width\\
\hline
Total & 729 c^2+131 c s+663s^2 \\
 K_1(1270)\pi & 15.6 c^2-9.12 c s+326s^2 \\
 K\rho & 153 c^2+165 c s+44.7 s^2 \\
 K^{*} \pi  & 155 c^2+160 c s+41.8 s^2 \\
 K\eta& 77.4 c^2-136 c s+60.0 s^2 \\
 K^{*}\rho & 132 c^2-34.3 c s+6.33 s^2 \\
 K \pi & 25.4 c^2-83.9 c s+692 s^2 \\
 K \phi & 74.4 c^2+52.7 c s+9.35 s^2 \\
 K h_1& 1.33 c^2-18.1 c s+78.0 s^2 \\
 K \omega& 51.0 c^2+55.0 c s+14.8 s^2 \\
 K^{*} \omega & 35.4 c^2-9.19 c s+1.69 s^2 \\
 K_1(1400)\pi & 3.63 c^2-10.2 c s+7.71 s^2 \\
 K\text{$\eta $'} & 2.64 c^2-4.81 c s+2.19 s^2 \\
 K_2^*(1430) \pi& 1.99 c^2+1.85 c s+0.429  s^2 \\
 %K^{*}\eta & 0.972 c^2+0.786 c s+0.159  s^2 \\
\hline
\hline
\label{tab:k1680}
\end{array}\]
\end{table}
\fi

As shown in Table \ref{tab:1D}, $K^*(1680)$ as a pure $1^3D_1$ state mainly decays into final states $K_1(1270)\pi$, $K h_1$, $K\pi$, and $K\eta$, while the $K^*\pi$ and ~$K\rho$  modes also have sizable contributions.
We notice that the obtained ratios of partial decay widths of $K\pi$, $K^*\pi$, and $K\rho$ in this work are comparable with experimental data given in PDG.
Since the branching ratios of the $K\pi$, $K^*\pi$, and $K\rho$ decay channels given by PDG are $38.7\%$, {$29.9\%$}, and {$31.4\%$}, we conclude that the remaining $K_1(1270)\pi$ decay channel has a
very small width. However, our calculation shows that $K_1(1270)\pi$ is a main contribution to the total width which is consistent with conclusion from the former analysis in \cite{Barnes:2002mu} but contradicts with the present experimental data. Here we and the authors of Ref. \cite{Barnes:2002mu} adopted the mixing angle $\theta_{1P}=45^\circ$ \cite{Blundell:1996as,Ebert:2009ub} in the corresponding calculations. It is obvious that we need to face this puzzle in this channel.
More experimental and theoretical efforts are needed to clarify this point. %Later, we will discuss the angle $\theta_{1P} $ later.

$K_2(1770)$ and  $K_2(1820)$ satisfy
\begin{equation}
{
\left( \begin{array}{c} | K_2(1770)\rangle \\ |K_2(1820)\rangle  \end{array} \right) \approx
\left( \begin{array}{cc} \cos{\theta_{1D}} & \sin{\theta_{1D}} \\
                         -\sin{\theta_{1D}} & \cos{\theta_{1D}} \end{array} \right)
\left( \begin{array}{c} |1^1D_2\rangle \\ |1^3D_2\rangle  \end{array} \right).}
\label{kmixing}
\end{equation}
\begin{figure}[htbp]
\centering%
\begin{tabular}{c}
\scalebox{0.50}{\includegraphics{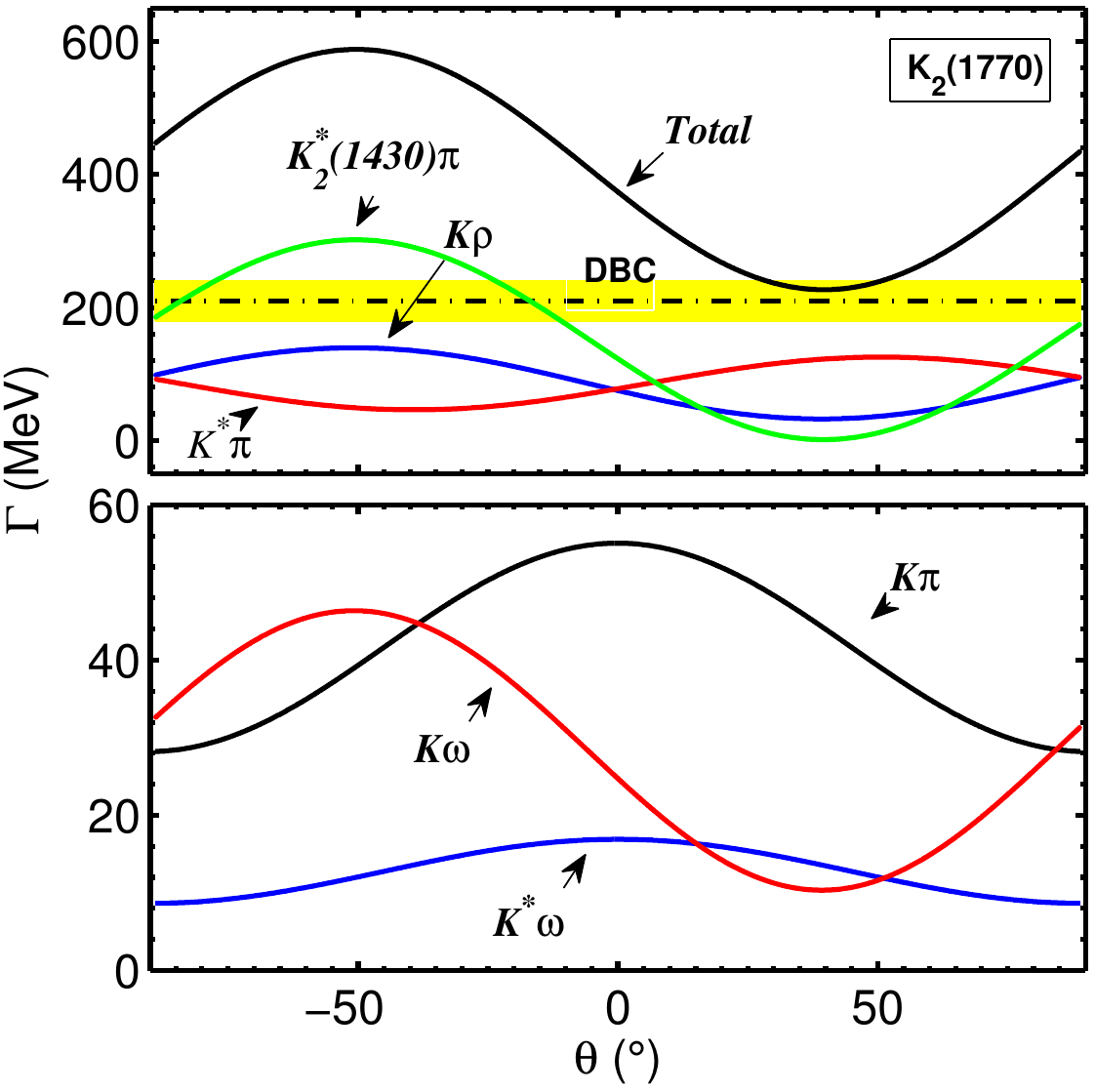}}
\end{tabular}
\caption{$\theta_{1D}$ dependence of the width of $K_2(1770)$, where the dot-dashed line is the experimental value of DBC \cite{Firestone:1972st}.}
 \label{fig:K21770}
\end{figure}
\par
{According to Fig. \ref{fig:K21770}, we find that $K_2(1770)$ mainly decays to $K_2^*(1430)\pi$, $K^*\pi$, $K\pi$, and $K\omega$.
Experiments show that  $K_2^*(1430)\pi$ is the dominant decay  mode of  $K_2(1770)$ \cite{Agashe:2014kda} which indicates that $\theta_{1D}$ favors the value less than zero.}
 \par
$K_3^*(1780)$ together with $\rho_3(1690)$, $\omega_3(1670)$, and $\phi_3(1850)$ forms a $1^3D_3$ nonet. We give its mass 1.781 GeV by the {MGI} model, which is consistent with experiment data 1.776 GeV.
As shown in Table \ref{tab:1D}, even though $K^*\rho$ is the dominant decay mode of $K_3^*(1780)$, it is not observed in experiments so far.
The channel $K^*\omega$ has sizable contribution to its total decay width, which is still missing in experiment.
 A final state $K\pi$ largely contributes to the total width and theory and experiments are consistent to each other.
 The branching ratio ${\Gamma_{K^*\pi}}/{\Gamma_{K\pi}}$ agrees with experimental data~\cite{Baubillier:1984wi}.
 \begin{table}[htbp]
 \renewcommand{\arraystretch}{1.3}
\caption{The decay widths of  $1^3 {D}_1$ and $ 1^3 {D}_3$ states. Here, $K^*(1680)$ and $K_3^*(1780)$ are assigned to be the pure $1^3 {D}_1$ state  and $ 1^3 {D}_3$ states, respectively. The unit of the width is MeV.  \label{tab:1D}}
\tabcolsep=1pt
\begin{center}
\begin{tabular}{cccccccccccc}
\hline\hline
 {State} &Channel&This work&Ref.~\cite{Barnes:2002mu}&Experiment \\
 \hline

 $1^3 {D}_1$
                             & $K\pi$ & 69.2&45& -- \\
                             & $K^*\pi $ &41.8&25& -- \\
                              &$K\rho $&44.7 &26&--\\
                              &$K\eta $ &64.4 &53&--\\
                              &$K^* \rho$&6.33 &2&--\\

                             % &$K(1460)\pi$&1.8 &1.68[2.89]&--\\

                              &$K^*\omega$ &1.69&1&--\\

                              &$K h_1$ &78&33&--\\
                               &$K_1(1270)\pi$ &330&145&--\\
                                &$K_1(1400)\pi$ &7.86&0&--\\
                                  &$K \phi$ &9.35&45&--\\

                              &Total & 653&348&$426\pm18\pm30$~\cite{Aston:1987ir} \\

                                                    &$\Gamma_{K\pi}/\Gamma_{K^*\pi}$&1.66 &1.8&$2.8\pm1.1$~\cite{Aston:1984ab}\\
                                                     &$\Gamma_{K\rho}/\Gamma_{K\pi}$ &0.65&0.58&$1.2\pm 0.4$~\cite{Aston:1984ab}\\
                                                     &$\Gamma_{K\rho}/\Gamma_{K^*\pi}$&1.07&1.04&$1.05^{+0.27}_{-0.11}$\\
                                                     \hline

$1^3 {D}_3$ &$K^*\rho$ &118&42 &--\\
                         & $K\rho$ & 20.1&10&$74\pm10$~\cite{Blundell:1996as} \\
                         & $K^*\omega $&36.4 &12& -- \\
                        &$K\pi $&38.1 &40&$31.7\pm3.7$~\cite{Blundell:1996as}\\
                         &$K^*\pi $ &28.5&14 &$45\pm 7$~\cite{Blundell:1996as}\\
                          &$K\omega$ &6.45&3&--\\
                          &$K\eta$ &9.67&19 &$48\pm 21$~\cite{Agashe:2014kda},$15\pm 6$~\cite{Aston:1987ey}\\

                           &$K_1(1270)\pi$ &1.68&1 &--\\

                           &$K_1(1400)\pi$ &2.80&1 &--\\
                             &$K_2^*(1430)\pi$&4.18&1 &$<25$~\cite{Aston:1987ey}\\

                         &Total & 266&145&$225\pm60$~\cite{Aston:1980bp} \\
                                                    &$\Gamma_{K\rho}/\Gamma_{K^*\pi}$ &0.702 &0.71&$1.52\pm0.23$~\cite{Baubillier:1984wi}\\
                                                     &$\Gamma_{K^*\pi}/\Gamma_{K\pi}$ &0.748 &0.35&$1.09\pm 0.26$~\cite{Baubillier:1984wi}\\
                                                     &$\Gamma_{K\eta}/\Gamma_{K\pi}$ &0.253 &0.48&$1.6\pm0.7$~\cite{Agashe:2014kda}\\
                                                     &$\Gamma_{K\pi}/\Gamma_{Total}$ &0.143 &0.28&$0.188\pm0.010$ \cite{Agashe:2014kda}\\

                                                     &$\Gamma_{K\rho}/\Gamma_{Total}$&7.5\%&6.9\%&{$(31\pm9)\%$} \cite{Agashe:2014kda}\\

                                                      &$\Gamma_{K^*\pi}/\Gamma_{Total}$ &10.7\%&9.7\%&{$(20\pm5)\%$} \cite{Agashe:2014kda}\\
                                                       &$\Gamma_{K\eta}/\Gamma_{Total}$ &3.6 \%&13\%&{$(30\pm13)\%$} \cite{Agashe:2014kda}\\
\hline
\hline
\end{tabular}
\end{center}
\end{table}
 \par
 Next, let us focus on $K_2(1820)$.
 According to Fig \ref{fig:K21820}, one notices that $K_2(1820)$ probably decays to $K\pi\pi$, $K_2^*(1430)\pi$, $K f_2(1270)$, $K^*\pi$, and $K\omega$, in which $K\pi\pi$ comes from $K\rho$  channel.
 As seen from $\theta_{1D}$ dependence of the widths of
 $K_2(1820)$ in Fig. \ref{fig:K21820}, we notice that contributions of $K\rho$ and $K f_2(1270)$ are large when $\theta_{1D}<0$, which indicates that it is very likely that $\theta_{1D}$ is smaller than zero that is consistent with the previous analysis for  $K_2(1770)$.
Because of absence of the experimental information, we cannot confirm the angle $\theta_{1D}$, while our results will be helpful for the future experiments to study this state.
\begin{figure}[htbp]
\centering%
\begin{tabular}{c}
\scalebox{0.35}{\includegraphics{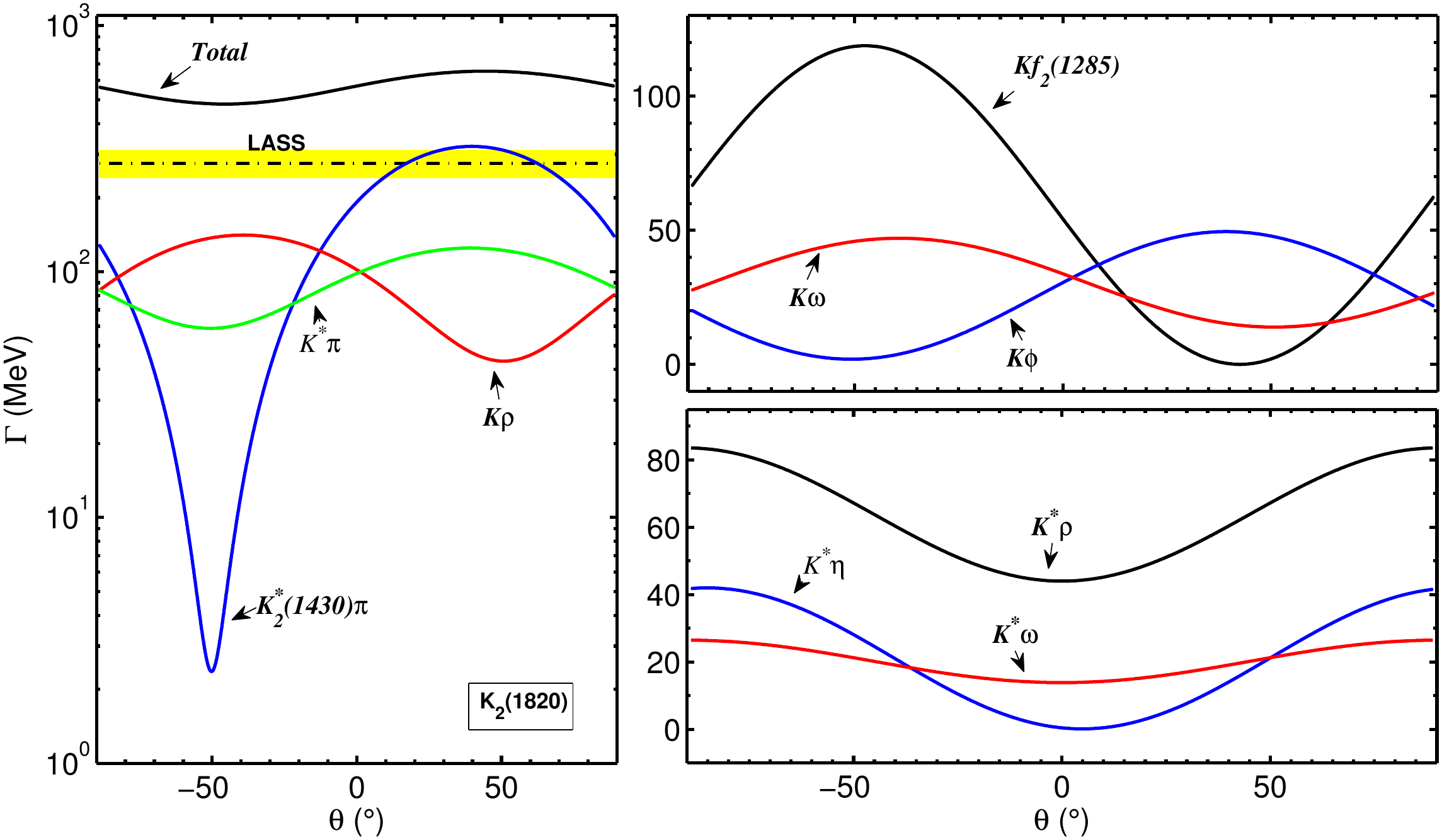}}
\end{tabular}
\caption{$\theta_{1D}$ dependence of the  decay width of $K_2(1820)$.
Here a dot-dashed line LASS with a yeallow band is taken from \cite{Aston:1986jb}.}
 \label{fig:K21820}
\end{figure}

\subsubsection{2$D$ states}

As one of the 2$D$ states is missing, using the familiar program with 2$P$ states, we obtain
\begin{equation}
\left( \begin{array}{c}  |K_2(1990)\rangle \\ |K_2(2250)\rangle \end{array} \right) \approx
\left( \begin{array}{cc} \cos{\theta_{2D}} & \sin{\theta_{2D}} \\
                         -\sin{\theta_{2D}} & \cos{\theta_{2D}} \end{array} \right)
\left( \begin{array}{c} |2^1D_2\rangle\\ |2^3D_2\rangle \end{array} \right),
\label{kmixing}
\end{equation}
where $K_2(1990)$ is obtained from an equation similar to Eq.~(\ref{mpmpp}).

According to Table \ref{tab:k22250}, one finds that $ K_2(2250)$ as a 2$D^\prime$ state mainly decays into $ K a_2{(1700)}$,
 $K_2^*(1980)\pi  $, $ K_3^*{(1780)} \pi$, $ K^*{(1410)} \pi $,  and $K_2^* \pi$.
  $K_2^*\pi$ and $K f_2(1270)$ have been observed in experiments which have sizable contributions to the total width.
 Besides, $K\rho$, which is an important decay channel in our result, can decay into $K\pi\pi$ that is observed in experiment.
 On the other hand, the theoretical total width is larger than the experimental value 180 MeV given in PDG.
 We need more experimental information to study this 2$D^\prime$ state, to test our results, and to have more detailed decay widths to ascertain the value of $\theta_{2D}$.

\begin{table*}[htbp]
\caption{Strong decay information of $K_2(2250)(2D^\prime)$, where $s$ and $c$ represent sine and cosine functions. The unit of the width is MeV. \label{tab:k22250}}
\[\begin{array}{cccc}
\hline
\hline
\text{Decay channel}&\text{Width}&\text{Decay channel}&\text{Width}\\
\hline
 \text{Total} &956 c^2-267c s+ 1044 s^2 & K a_2{(1700)} & 115 c^2-188c s +77.1 s^2\\
    K_2^*(1980) \pi& 82.6 c^2-136 c s +57.4 s^2& K_3^*{(1780)} \pi  & 67.3 c^2-135 c s +71.4 s^2\\
 K^*{(1410)} \pi  & 75.5 s^2+55.6 c^2-97.2 c s & K_2^* \pi  &58.9 c^2-70.9 c s+ 64.2 s^2\\
 K a_2 &52.9 c^2-57 c s+ 55.1 s^2 & K \rho _3{(1690)} &33.7 c^2-64.5 c s+ 31 s^2 \\
% K_1(1650)\pi & 26.7 c^2-59.2 c s +32.8 s^2& K_2^* \rho  & 11.4 c^2-37.7 c s +45.4 s^2\\
 K^* \pi  &31.3 c^2-50.3 c s+ 21.1 s^2 & K \rho(1450) &42.4 c^2-23.3 c s + 37.6 s^2\\
 K \rho  &29.7 c^2-48c s+ 19.9 s^2 &  K^*a_2 &9.24 c^2-30.9 c s + 37.9 s^2\\
 K^*{(1410)} \rho  &22.1 c^2+ 42.1 s^2 & K^* {\pi(1300)} &25.6 c^2-40.4 c s + 17.3 s^2\\
 K^* \rho  & 37.7 c^2+40.2 s^2 & K_1{(1400)} \rho  &15.5 c^2-42.8 c s+ 16.7 s^2 \\
 %K_1 \rho  &15.5 c^2-42.8 c s+ 16.7 s^2 & K^*{(1410)} \eta  &0.381 c^2-2.11 c s+ 37s^2 \\
 K \omega _3{(1670)} &15.1 c^2-29c s+ 14s^2 & K f_2 &19.9 c^2-18.1 c s+ 20.1 s^2 \\
 K f'_2{(1525)} &19.9 c^2-18.1 c s+ 20.1 s^2 & K_2^* \omega  &3.49 c^2-12.9 c s+ 15.6 s^2 \\
 K \omega  &9.85 c^2-15.9 c s + 6.61 s^2&  K^* b_1& 14.3 c^2-0.4 c s +16.1 s^2\\
 K^*a_1  &6.74 c^2-7c s+ 14.1 s^2 & K a_0{(1450)} &4.36 c^2-13.7 c s+ 10.8 s^2 \\
 K {\omega(1420)} &14.1 c^2-2.21 c s+  13.6 s^2& K^* \eta  &0.101 c^2-0.517 c s+ 14.2 s^2 \\
 K^* \omega  &12.5 c^2 + 13.2 s^2 & K^*{(1410)} \omega  &6.37 c^2 + 12.3 s^2\\
 K^* h_1 & 10.2 c^2-3.25 c s+10.6 s^2 & K^*  f_2&6.31 c^2-7.46 c s+ 9.34 s^2 \\
 K^*{(1680)} \pi  &4.98 c^2-11c s+ 6.44 s^2 & K^* {\eta(1295)} &0.0195 c^2-0.914 c s+ 10.7 s^2 \\
 K^* \phi  &8.35 c^2+ 6.81 s^2 & K_1{(1400)} \omega  & 1.26 c^2-1.09 c s +8.18 s^2\\
 %K_1 \omega  &1.26 c^2-1.09 c s + 8.18 s^2&  K(1460) \rho & 4.7 c^2-7.67 c s +3.14 s^2\\
 \hline
 \hline
\end{array}\]
\end{table*}

We use 1994 MeV as the mass of the partner of $K_2(2250)$ with $\theta_{2D} \approx \pm44^\circ$, and calculate the strong decay of this state as shown in Table \ref{tab:k21990}. According to this Table, we can find that its main decay channels are  $K^*\pi$, $K\rho$, $K_2^*\pi$, and $K_3^*(1780)\pi$.\\

 \begin{table}[htbp]
 \renewcommand{\arraystretch}{1.2}
\caption{The main strong decay widths of $K_2(1990)$ as an $2D$ state, where $s$ and $c$ represent sine and cosine functions.
The unit of the width is MeV. \label{tab:k21990}}
%\begin{center}
 %\resizebox{!}{5.5cm}{
\[\begin{array}{cc}
\hline\hline
\text{Decay channel}&\text{Width}\\
 \hline
 \text{Total} & 97.9 c^2-17.3cs+120s^2 \\
 K^* \pi  & 15.1c^2-22cs+19.6s^2 \\
 \text{K} \rho  & 12.5c^2+19.2 cs+16.4s^2 \\
 K_2{}^* \pi  & 17.9c^2-16.5cs+7.59s^2 \\
 K_3^*\text{(1780)} \pi  & 7.16c^2-14.4cs+7.26s^2 \\
  \hline
   \hline
\end{array}\]
%\end{center}
 \end{table}

\subsection{$F$-wave kaons}

\subsubsection{$1F$ states}
%\paragraph{$K_2^*(1980)$ is $1^3F_2$ or $2^3P_2$ state? }
In this subsection, we discuss possibility of different assignments of $K_2^*(1980)$ from two aspects, mass and decay information.
%\subparagraph{Mass analysis}
%\subparagraph{As $1^3F_2$ state}
In 1987, LASS reported a structure in the reaction $K^- p \to \bar K^0 \pi^+ \pi^- n$ %, which probably includes the kaons with $1^3F_2$ or $2^3P_2$ quantum numbers in the final state
\cite{Aston:1986jb}, and they obtained resonance parameters $M=(1973\pm8\pm25)$ MeV and $\Gamma=(373\pm33\pm60)$ MeV. This is the particle  called $K_2^*(1980)$ listed in PDG \cite{Agashe:2014kda}.
Barnes {\it et al.}~\cite{Barnes:2002mu} have the viewpoint that $K_2^*(1980)$ is a $1^3F_2$ state, and give the total width 300 MeV. On the other hand,
our results show that  the mass of a $1^3F_2$ state is about 2093 MeV. Ebert {\it et al.}~\cite{Ebert:2009ub} predict $1^3F_2$ state with the mass 1964 MeV. As the partner of an iso-vector of $1^3F_2$, $a_2(2030)$ is well established in Ref. \cite{Pang:2014laa}. {In the same nonet, the meson which contains one $s$ quark is probably heavier than the mesons which only contain $u/d$ quarks.}
{Along this line,} the mass of $1^3F_2$ state of the kaon should be lager than 2030 MeV, so that the mass $K_2^*(1980)$ is a bit small as a $1^3F_2$ state.
%\subparagraph{As $2^3P_2$ state}
The papers of Refs. \cite{Godfrey:1985xj,Ebert:2009ub,Barnes:2002mu} and this work give the mass for a $2^3P_2$ state
 1938, 1896, 1850, and 1870 MeV, respectively, and for its iso-vector partner is $a_2(1700)$ \cite{Pang:2014laa}, the mass of $K_2^*(1980)$ is  a bit larger as a $2^3P_2$ state.
% \par
We should, of course, combine the decay information of $K_2^*(1980)$ to determine which possibility of its assignment we should take.
%\subparagraph{Decay information}

%\subparagraph{As~$1^3F_2$ state}
Both Ref. \cite{Barnes:2002mu} and this work (see Table \ref{tab:1F2}) show that $K_1(1270)\pi$ is the dominant decay channel when we treat $K_2^*(1980)$ as a $1^3F_2$ state, even though the channel is not observed in experiments.
$K_2(1770)\pi$, $Kb_1$, $Ka_1$, {$K\pi$, $K\rho$,} and $K^*\pi$ modes, among which $K\rho$ and $K^*\pi$ have been reported in the experiment \cite{Agashe:2014kda}, also have sizable contributions, where we take $\theta_{1D}=-39^\circ$. If $K_2^*(1980)$ is a $1^3F_2$ state,
our prediction for the channels {$K_1(1270)\pi$,} $K_2(1770)\pi$, $Kb_1$, $Ka_1$, and $K\pi$ will be helpful for the
experiment to test this assignment.

%\subparagraph{As $2^3P_2$ state}

Both  Ref. \cite{Barnes:2002mu} and our results show that $K_2^*(1980)$ is $2^3P_2$ state, where Ref. \cite{Barnes:2002mu} takes the mass 1850 MeV and we take the experimental value  1973 MeV. For this reason, the results between theirs and this work have some difference. The main decay modes are $K^*\rho$, $K\pi$, $K^*\pi$, $K\rho$, $K^*\eta$, $K\eta^\prime$, and $K^*\omega$.
Besides the $K\rho$ and $K^*\pi$, one notices that $Kf_2$ has been observed  in experiments which has sizable contributions in theory.
Hence $K_2^*(1980)$ assigned to $2^3P_2$ is also reasonable.
%\subparagraph{Conclusion}
\par
Finally, let us draw a rough conclusion for $K_2^*(1980)$.
According to the mass analysis, the mass of $K_2^*(1980)$ is a bit small when assigned to a $1^3F_2$ state and a bit large when assigned to a $2^3P_2$ state.
According to the decay information, $K_2^*(1980)$ is in favor of a $2^3P_2$ state. We still, however, need more experimental information to test our assignment for $K_2^*(1980)$.
What is more important is that  we give the prediction that the partial widths of
$K_2^*(1425)\pi$, $K^*(1410)\pi$,
$K^*\omega$, $K^*\rho$, and $K^*\eta$  treating $K_2^*(1980)$ as $1^3F_2$ will be much larger than  those of the case of a $2^3P_2$ state.
The experimental study of these decay modes combined with our prediction will help us determine assignment of $K_2^*(1980)$.
Besides the above, our prediction can help the future experiments find the missing$1^3F_2$ or $2^3P_2$ state.
%\paragraph{$K_4^*(2045)$ as $1^3F_4$ state }
According to Table \ref{tab:kmassground1}, one can notice that our spectral results are consistent with the mass of $K_4^*(2045)$ given by PDG when we treat it as $1^3F_4$ state.
As for the strong decay of $K_4^*(2045)$, one can notice that both the results of Ref. \cite{Barnes:2002mu} and this work (Table \ref{tab:1F4})
  show that $K^*\rho$, $K\pi$, $K^*\pi$, $K^*\omega$, and $K\rho$ are the main decay channels.
 The PDG gives two partial width ratios: one is
 ${\Gamma_{K\pi}}/{\Gamma_{Total}}=(9.9\pm1.2)\%$ and our result is ${\Gamma_{K\pi}}/{\Gamma_{Total}}=8.4\%$ which is consistent with the experiment. Another is ${\Gamma_{K^*\phi}}/{\Gamma_{Total}}=(1.4\pm0.7)\%$ and we obtain ${\Gamma_{K^*\phi}}/{\Gamma_{Total}}=1.54\%$, which is consistent with the experiment as well. On the other hand, Ref. \cite{Barnes:2002mu} obtained $21\%$ and $3.1\%$ for these two partial width ratios, which are different from the experiment. these results, of course, prove the superiority of the accurate meson wave functions we have obtained.

\begin{table}[htbp]
\renewcommand{\arraystretch}{1.3}

\caption{The strong decay widths of $K_4^*(2045)$ assigned to a $1^3F_4$ state. The unit of the width is MeV. \label{tab:1F4}}
\begin{center}
% \resizebox{!}{3.5cm}{
\begin{tabular}{cccccccccccc}
\hline\hline
 {State} &Channel&This work&Ref.~\cite{Barnes:2002mu}&Experiment\\
 \hline
 $K_4^*(2045)$( $1^3 {F}_4 $)
       &$K^*\rho$ &84.9 &29&--\\
           & $K\rho $& 16.1&7& -- \\
         & $K^*\omega $ & 27.7&9& -- \\
             &$K\omega$ &5.24 &2&--\\
           &$K\pi $ &21.0 &21&--\\
           &$K^*\pi$ &20.5 &8&--\\
             &$K^*(1410)\pi$ &2.91&0 &--\\

              &$K_1(1270)\pi$ &11.4&2&--\\
                &$K_1(1400)\pi$ &6.47&2&--\\
&$K\phi $&0.783 &1&--\\

&$K a_1 $&4.17&1 &--\\
&$K a_2 $&13.5 &1&--\\
&$K b_1 $ &13.0 &2&--\\
&$K^* \phi $ &3.84&3 &--\\
&$K_2^*(1430) \pi $ &15.9&2&--\\
   &$K_1(1270)\eta$ &3.13&1&--\\

%$K_1 \pi$& $&-- &13.5 &--\\
%&$K b_1 $&-- &13.0 &--\\

     &Total& 250&98&$198\pm30$ \\

      &$\Gamma_{K\pi}/\Gamma_{Total}$ &8.40\%&21\%&$9.9\pm1.2\%$~\cite{Aston:1987ir}\\
        &$\Gamma_{K^*\phi}/\Gamma_{Total}$&1.54\%&3.1\% &$1.4\pm0.7\%$~\cite{Aston:1986rm}\\

\hline
\hline
\end{tabular}
\end{center}
\end{table}
\par

Papers of Refs.~\cite{Godfrey:1985xj,Ebert:2009ub,Barnes:2002mu} and this work give the mass of a $1F$ state
$2131$, $2009$, $2050$ , and $2075$ MeV {(which we call $K_3(2075)$, and  strictly speaking, this state is a pure $1^1F_3$ state, here we assume the physical state $1F$ has this mass)}, respectively, among which one notices that the last two results are almost identical. $K_3(2075)$ is assigned to the missing $1F$ state.
We present the $\theta_{nF}$ dependence of widths for these two cases in Tables~\ref{tab:k323241}.
The total width of a $1F$ state with the mass 2075 MeV is about
    {$(400-600)$} MeV, which means that the predicted $K_3(2075)$ is a broad state and it is not
easy to identify $K_3(2075)$ in experiments.
    {Its main decay channels are $K_3^*{(1780)} \pi $, $K^*\rho$,  $ {K^*} \pi$, $ K a_2$, $ K\rho$ and $K_2^*\pi$}.
\subsubsection{$2F$ states}
 $K_3(2F)$ and $K_3(2F^{\prime})$  mixing satisfies
\begin{equation}
\left( \begin{array}{c}  |K_3(2320)\rangle \\ |K_3(2360)\rangle \end{array} \right) \approx
\left( \begin{array}{cc} \cos{\theta_{2F}} & \sin{\theta_{2F}} \\
                         -\sin{\theta_{2F}} & \cos{\theta_{2F}} \end{array} \right)
\left( \begin{array}{c} |2^1F_3\rangle\\ |2^3F_3\rangle \end{array} \right),
\label{kmixing}
\end{equation}
where $K_3(2360)$  is obtained from an equation similar to Eq. (\ref{mpmpp}).
The total width of $K_3(2320)$ is nearly $(180-200)$ MeV which is consist with the data of OMEG  {$(150\pm 30)$} MeV~\cite{Armstrong:1983gt}.  $K_3(2320)$ mainly decays
to $K_3^*(1780)\pi$, $K^*(1410)\pi$, $K\rho$ and $K^*\pi$.
The total width of $K_3(2360)$ is nearly $(80-120)$ MeV.  $K_3(2360)$ mainly decays
to $K_3^*(1780)\pi$, $K\rho_3(1690)$, $K\rho$ and $K^*\pi$ which are given in Table \ref{tab:k32360}. Although we cannot give the mixing angle of these two states for the lack of experimental information, our theoretical results can be helpful for studying these two states in the future experiments.

  \setlength\arraycolsep{1.4pt}%
\begin{table*}[htbp]
\caption{Strong decay information of $K_3(2320)$ as an $2F$ state and (predicted) $K_3(2075)(1F)$ depending on their mixing angle, where $s$ and $c$ represent sine and cosine functions. The unit of the width is MeV. \label{tab:k323241}}
\[\begin{array}{ccccc}
\hline
\hline
\text{Decay channel}&K_3(2075) \text{~as~1$F$ state}&K_3(2320) \text{~as~2$F$ state}\\
\hline
 \text{Total} & 464c^2+182  cs +530s^2& 189c^2+21.2 cs+202s^2 \\
 K_3^*{(1780)} \pi  & 107c^2+245 cs+141s^2 & 24c^2+42.4 cs+21.3 s^2 \\
 K^*{(1410)} \pi   & 6.93 c^2+0.975 cs+7.07 s^2 & 20.3 c^2+27.2 cs+16.4 s^2 \\
 K \rho   & 38.7 c^2+11.1 cs +40.3 s^2& 11c^2+20.4 cs+13.9 s^2 \\
 K^* \pi  & 40.5 c^2+4.03 cs+39.9 s^2 & 10.4 c^2+20.2 cs+13.3 s^2 \\
 K a_2{(1700)} &0 & 12.9 c^2+13.7 cs+9.98 s^2 \\
  K_2^*(1980)\pi  & 0 & 11.5 c^2+12.3 cs +8.81 s^2\\
 K \rho _3{(1690)} & 0 & 8.44 c^2+16.2 cs +8.19 s^2\\
%\pi  K_1(1650) & 2.59 s^2 & 16.3 s^2 \\
 K^*{(1410)} \rho & 0 & 13.5 c^2+9.77 s^2 \\
 K^* \rho  & 55.7 c^2+43.7 s^2 & 12.3 c^2+8.63 s^2 \\
 K {\rho (1450)}& 1.51 c^2+3.33 cs +1.99 s^2& 7.87 c^2+1.95 cs +8.15 s^2\\
 K \omega _3{(1670)} & 0 & 4.04 c^2+7.94 cs +4.13 s^2\\
 K^*{(1410)} \eta   & 0.0642 c^2+0.408 cs +0.703 s^2& 0.163 c^2+0.557 cs +7.7 s^2\\
 K \omega & 12.8 c^2+4.01 cs +13.3 s^2& 3.63 c^2+6.79 cs +4.61 s^2\\
 K^* {\pi (1300)}& 0 & 3.56 c^2+6.55 cs +4.51 s^2\\
 K_4^*{(2045)} \pi   & 0 & 4.3 c^2+4.62 cs+1.56 s^2 \\
  K(1460)\rho & 0 & 2.41 c^2+5.2 cs+3.16 s^2 \\
 K^*{(1680)} \pi  & 0.248 c^2+1.02 cs+1.05 s^2 & 3.33 c^2+4.79 cs +1.73 s^2\\
 K_2^* \rho   & 0 & 3.96 c^2+1.4 cs +3.94 s^2\\
 K^*{(1410)} \omega  & 0 & 4.14 c^2+2.97 s^2 \\
 K^* \omega  & 18.1 c^2+14.2 s^2 & 4c^2+2.83 s^2 \\
 K a_0{(1450)} & 1.15 c^2 & 3.91 c^2 \\
% K_1 \pi & 30.8 s^2 & 3.67 s^2 \\
  K^*a_2 & 0 & 2.73 c^2+0.957 cs+2.7 s^2 \\
 K {\omega (1420)} & 0.93 c^2+1.94 cs+1.21 s^2 & 2.89 c^2+0.472 cs+2.82 s^2 \\
 K^* \eta  & 0.106 c^2+1.54 cs +15.4 s^2& 0.0206 c^2+0.429 cs+2.86 s^2 \\
 K a_2 & 50.3 c^2+54.7 cs+37.3 s^2 & 1.84 c^2+2.13 cs+0.972 s^2 \\
 K a_1 & 24.5 s^2 & 2.33 s^2 \\
 K b_1 & 6.76 s^2 & 2.07 s^2 \\
 K_0^*(1945)  \pi & 0 & 1.66 c^2 \\
 % K(1460) \omega  & 0 & 0.688 c^2+1.51 cs +0.905 s^2\\
K \phi & 6.29 c^2+10.4 cs+7.78 s^2 & 0.787 c^2+1.33 cs+0.98 s^2 \\
 K_2^* \omega & 0 & 1.28 c^2+0.452 cs+1.28 s^2 \\
 K f_2 & 20c^2+14.6 s^2+21.9 cs & 1.03 c^2+1.05 cs +0.123 s^2\\
 %K f'_2{(1525)} & 25.2 c^2+36.1 cs+31.3 s^2 & 20c^2+21.9 cs+14.6 s^2 & 1.03 c^2+1.05 cs+0.123 s^2 \\
  %f_2 K^* & 0 & 0.916 c^2+0.311 cs +0.811 s^2\\
 K_2^* \pi  & 55.5 c^2+60.8 cs+40.4 s^2 & 0.742 c^2+0.853 cs+0.366 s^2 \\
 K_2^* \eta  & 8.55 c^2+1.11 cs+0.0976 s^2 & 0.773 c^2+0.0199 cs+0.000205 s^2 \\
 K^* \phi  & 3.01 c^2+2.04 s^2 & 0.474 c^2+0.585 s^2 \\
  %K_1{(1400)} \rho & 0 & 0.555 c^2+0.000734 cs+0.359 s^2 \\
 K^* {\eta '} & 2.7 c^2+1.74 cs+0.309 s^2 & 0.429 c^2+0.206 cs+0.0584 s^2 \\
  %\eta  K_1(1650)& 0 & 0.434 s^2 \\
 % \pi  K_2((1770)& 0.312 c^2+0.701 cs +0.394 s^2& 0.344 c^2+0.330 cs+0.0908 s^2 \\
 K_1(1270) \rho & 5.65 c^2+5.95 cs+5.16 s^2 & 0.263 c^2+0.216 cs+0.268 s^2 \\
 K f_1{(1420)} & 2.93 s^2 & 0.0280 c^2+0.366 s^2 \\
 K h_1& 2.89 s^2 & 0.198 c^2+0.347 s^2 \\
 K f_1  & 6.98 s^2 & 0.136 c^2+0.225 s^2 \\
  %K_1{(1400)} \omega & 0 & 0.184 c^2+0.000247 cs+0.119 s^2 \\
  %K \pi _2 & 0 & 0.158 s^2 \\
  %b_1 K^* & 0 & 0.125 c^2+0.0586 cs+0.131 s^2 \\
 % K_0^*{(1430)} \pi  & 0.150 c^2 \\
  %h_1 K^*& 0.567 c^2+0.506 cs+0.610 s^2 & 0.0295 c^2+0.00467 cs+0.147 s^2 \\
 % K_1(1270) \omega & 1.20 c^2+1.09 s^2+1.26 cs & 0.101 c^2+0.0869 cs +0.102 s^2\\
  % %K^*{(1680)} \eta  & 0 & 0.00362 c^2+0.0448 cs+0.14 s^2 \\
 % K_1(1270) \phi  & 0 & 0.0911 c^2+0.0272 cs+0.0903 s^2 \\
 % f_1 K^* & 0 & 0.0235 c^2+0.0114 cs+0.0549 s^2 \\
 % a_1 K^* & 0.0492 c^2+0.000115 cs +0.0398 s^2 \\
 % K_1{(1400)} \pi   & 6.08 s^2 & 0.0472 s^2 \\
  %K_0^*{(1430)} \rho  & 0 & 0.046 s^2 \\
 % K_1(1270) \eta& 1.06 s^2 & 0.0182 s^2 \\
 % K_0^*{(1430)} \omega& 0 & 0.0154 s^2 \\
  %K^* {\omega (1420)}& 0 & 0.00934 c^2+0.00623 s^2 \\
  %K_0^*{(1430)} \eta & 1.03 c^2 & 0.00866 c^2 \\
 \hline
 \hline
  \end{array}\]
\end{table*}

\renewcommand{\arraystretch}{1.3}
\begin{table}[htbp]
\caption{The main strong decay widths of $K_3(2360)$ assigned to an  $2F^{\prime}$  state, where $s$ and $c$ represent sine and cosine functions. The unit of the width is MeV. \label{tab:k32360}}
\begin{center}
\[\begin{array}{cc}
\hline
\hline
 \text{Decay channel} &\text{Width} \\
 \hline
 \text{Total} & 93 c^2+34 cs+101s^2 \\
 K_3{}^*\text{(1780)} \pi  & 28.3 c^2+54.7 c s+30.4 s^2 \\
 \text{K} \rho _3\text{(1690)} & 16 c^2-22.8 c s+22.8 s^2 \\
 \text{K} \rho  & 15.1 c^2-20.3 c s+12.1 s^2 \\
 K^* \pi  & 14.3 c^2+20 c s+11.4 s^2 \\
 K_4{}^*\text{(2045)} \pi  & 2.52 c^2+11.9 c s+11.4 s^2 \\
 \text{K} \omega _3\text{(1670)} & 6.88 c^2-9.62 c s+9.62 s^2 \\
 K^* \rho  & 9.83 c^2 +14.3 s^2 \\
 \hline
 \hline
\end{array}\]
\end{center}
\end{table}

\subsection{$G$-wave kaons}

\subsubsection{$1G$ states}

$K_4(1G)$ and $K_4(1G^{\prime})$ mixing  satisfies

\begin{equation}
\left( \begin{array}{c}  |K_4(2310)\rangle \\ |K_4(1G^{\prime})\rangle \end{array} \right) \approx
\left( \begin{array}{cc} \cos{\theta_{1G}} & \sin{\theta_{1G}} \\
                         -\sin{\theta_{1G}} & \cos{\theta_{1G}} \end{array} \right)
\left( \begin{array}{c} |1^1G_4\rangle\\ |1^3G_4\rangle \end{array} \right),
\label{kmixing}
\end{equation}
We assume the mass of a $1G$ state is about ~$2309$ MeV, which we call $K_4(2310)$. The GI model~\cite{Godfrey:1985xj} and Ref.~\cite{Ebert:2009ub} give this mass
~$2422$ MeV and~$2255$ MeV, respectively, while~the mass of the $K_4$ state in PDG is~2490 MeV. According to Table~\ref{tab:kmassground1}, $K_4(2500)$ may be a $2G$ state. We predict the strong decay information of these two $G$
 wave states in Tables \ref{tab:k425001}.

\renewcommand{\arraystretch}{1.2}
 \begin{table*}[htbp]
\caption{The widths of (predicted) $K_4(2310)$ and $K_4(2500)$ depending on their mixing angle, where $s$ and $c$ represent sine and cosine functions. The unit of the width is MeV. \label{tab:k425001}}
\[\begin{array}{ccccc}
\hline
\hline
\text{Decay channels}&\text{Width ($K_4(2310)$ assigned to $1G$ state)}&\text{Width ($K_4(2500)$ assigned to~$2G$ state)}\\
\hline
 {Total} & 664 c^2+31c s+754 s^2 &222 c^2 + 7.5 c s + 247 s^2\\
  K_4^*(2045) \pi  & 90.2 c^2+201 c s+112 s^2 &14.6 c^2 + 28.7 c s + 14.6 s^2\\
 K_3^*(1780) \pi  & 80  c^2+168 c s+92.3 s^2 &21.6 c^2 + 44.1 c s + 24.7 s^2\\
  K \rho _3(1690) & 63.1 c^2+140. c s+77.6 s^2& 19  c^2 + 41.6 c s + 23.3 s^2\\
 K a_2 & 43.1 c^2+70.4 c s+40.9 s^2 &17.5 c^2 + 34. c s + 20.5 s^2\\
  K_2^* \pi  & 41.7 c^2+62.6 c s+38.6 s^2 &16.7 c^2 + 31.7 c s + 20.2 s^2\\
 K \omega _3{(1670)} & 24  c^2+52.8 c s+29.2 s^2 &7.11 c^2 + 15.5 c s + 8.69 s^2 \\
  K^* b_1& 22.2 c^2+45.2 c s+27.1 s^2&8.14 c^2 + 17.2 c s + 10. s^2 \\
  K^* \rho  & 29.9 c^2+28.2 s^2 &9.16 c^2 + 7.28 s^2\\
  K^* \pi  & 19  c^2+15  c s+17.3 s^2&7.05 c^2 + 3.28 c s + 7.42 s^2 \\
 K f_2 & 15.2 c^2+22.3 c s+13.3 s^2 &6.22 c^2 + 11.7 c s + 7.08 s^2\\
 %K f'_2(1525) & 3.69 c^2-9.51 c s+6.28 s^2 &6.22 c^2 + 11.7 c s + 7.08 s^2\\
 K^*a_1  & 10.9 c^2+22.5 c s+12.9 s^2 &4.18 c^2 + 8.74 c s + 5.1 s^2 \\
 K^* h_1 & 10.8 c^2+20.2 c s+12.9 s^2&3.74 c^2 + 7.88 c s + 4.6 s^2\\
 K \rho  & 17.5 c^2+7.35 c s+16.7 s^2 &7.76 c^2 + 4. c s + 8.21 s^2\\
   K^* a_2& 15.4 c^2+11.5 c s+9.12 s^2 &5.58 c^2 + 3.68 c s + 4.01 s^2\\
 K_2^* \rho  & 14.9 c^2+11.2 c s+8.83 s^2 &3.36 c^2 + 2.39 c s + 2.96 s^2\\
  K_2^* \eta  & 0.241 c^2+3.62 c s+17.1 s^2&0.139 c^2 + 0.96 c s + 2.35 s^2 \\
 K^*(1410) \pi  & 10.2 c^2+7.29 c s+9.41 s^2 &4.39 c^2 + 7.83 c s + 3.51 s^2 \\
 K a_1 & 3.07 c^2+11.1 c s+9.99 s^2 &1.19 c^2 + 2.63 c s + 1.46 s^2\\
  K^* f_2& 9.84 c^2+7.37 c s+5.98 s^2 &3.04 c^2 + 1.99 c s + 2.07 s^2\\
  K^* \omega  & 9.79 c^2+9.18 s^2&3.0 c^2 + 2.38 s^2 \\
 K \rho(1450) & 5.22 c^2+7.23 c s+6.02 s^2 &0.95 c^2 + 1.67 c s + 0.764 s^2 \\
 K \omega  & 5.77 c^2+2.3 c s+5.51 s^2&2.55 c^2 + 1.38 c s + 2.71 s^2 \\
 K^*{(1680)} \pi  & 2.31 c^2+6.41 c s+4.45 s^2 &0.439 c^2 + 0.2 c s + 0.0385 s^2\\
  % K_1(1650)\pi & 1.45 c^2+5.54 c s+5.31 s^2 &1.21 c^2 + 7.09 c s + 9.21 s^2\\
 K_2^* \omega  & 4.35 c^2+3.26 c s+2.57 s^2 &0.01 c^2 + 0.724 c s + 0.909 s^2\\
  K_0^*(1430) \pi  & 2.3 c^2+5.12 c s+2.84 s^2 &0.0247 c^2 + 0.232 c s + 0.546 s^2\\
   % K a_2{(1700)} & 2.3 c^2+5.1 c s+2.85 s^2 &6.34 c^2 + 10.1 c s + 5.79 s^2\\
     %K^* f_1& 2.07 c^2+4.43 c s+2.51 s^2&0.983 c^2 + 2. c s + 1.16 s^2 \\
  %K^* \eta  & 4.46 c^2+0.25 c s+0.0368 s^2 &0.814 c^2 + 0.199 c s + 0.013 s^2\\
  % K \omega(1420) & 2.24 c^2+2.51 c s+2.52 s^2 &0.329 c^2 + 0.586 c s + 0.263 s^2\\
 % K f_1 & 0.736 c^2+2.75 c s+2.56 s^2 &0.276 c^2 + 0.576 c s + 0.301 s^2\\
 %  K a_0{(1450)} & 2.01 c^2+3.1 c s+1.2 s^2 &1.36 c^2 + 0.0346 c s + 0.00022 s^2\\
  K_2^*(1980)\pi  & 1.43 c^2+3.14 c s+1.76 s^2 &9.02 c^2 + 16.7 c s + 9.55 s^2\\
 K \phi  & 1.92 c^2+2.01 c s+2.14 s^2&0.36 c^2 + 0.771 c s + 0.446 s^2 \\
K b_1 & 3.02 s^2 &0.988 s^2\\
  %K^* \pi (1300) & 0.964 c^2+2.06 c s+1.19 s^2 &0.671 c^2 + 0.708 c s + 0.592 s^2\\
 K^* \phi  & 2.08 c^2+1.67 s^2 &0.247 c^2 + 0.198 s^2 \\
   % K_2(1820)\pi & 1.27 c^2+4.29 c s+3.65 s^2& 0.347 c^2 + 0.6 c s + 0.464 s^2\\
   % K_2(1770)\pi & 1.32 c^2+1.47 c s+0.413 s^2 &0.158 c^2 + 0.0552 c s + 0.957 s^2\\
 % K^* \eta^{\prime} & 0.114 c^2+0.634 c s+1.46 s^2&0.031 c^2 + 0.209 c s + 0.351 s^2 \\
 K^*{(1410)} \eta  & 1.4 c^2+0.694 c s+0.209 s^2 &1.82 c^2 + 0.611 c s + 0.0524 s^2\\
 %  K h_1 & 1.38 s^2 &0.481 s^2\\
 K^*{(1410)} \rho  & 1.03 c^2+0.79 s^2 &6.55 c^2 + 6.31 s^2\\
 \hline
 \hline
\end{array}\]
\end{table*}

As shown in Table \ref{tab:k425001}, main decay modes of $K_4(2310)$ are {$K_4^*{(2045)}\pi $,
 $K_3^*{(1780)}\pi$, $K \rho _3{(1690)} $, $K a_2$ and~$  K_2^* \pi $ when $K_4(2310)$ is assigned to a $1G$ state. Its total width will be~$(710-880)$ MeV}, which is not easy to observe in experiments.

\subsubsection{$2 G$ states}
 $K_4(2500)$ and its partner  $K_4(2550)$ (predicted) satisfy
\begin{equation}
\left( \begin{array}{c}  |K_4(2500)\rangle \\ |K_4(2550)\rangle \end{array} \right) \approx
\left( \begin{array}{cc} \cos{\theta_{2G}} & \sin{\theta_{2G}} \\
                         -\sin{\theta_{2G}} & \cos{\theta_{2G}} \end{array} \right)
\left( \begin{array}{c} |2^1G_4\rangle\\ |2^3G_4\rangle \end{array} \right),
\label{kmixing}
\end{equation}
where $K_4(2550)$ is obtained from an equation similar to Eq. (\ref{mpmpp}).

 The total width of $K_4(2500)$ assigned to a $2G$ state is about~{$(230-290)$} MeV, which is consistent with the experimental value~$\sim$250 MeV \cite{Cleland:1980ya}. According to Table \ref{tab:k425001}, the main decay channels of $K_4(2500)$ are
$K_3^*(1780) \pi$, $K \rho _3(1690)$, $K a_2$, $ K_2^*\pi $,  $ K_4^*(2045)\pi $, $ K^*b_1$, $ K_2^*(1980)\pi $ and~$K \omega _3(1670)$, etc. {Information of these predicted decay widths is }important to study the mixing angle of this state for the future experiment.

The total width of $K_4(2550)$ assigned to a $2G^\prime$ state is about~{$(230-260)$} MeV. According to Table \ref{tab:k42550}, the main decay channels of $K_4(2550)$ are
$ K_3^*(1780)\pi $, $K \rho _3(1690)$, $K a_2$,  $ K_4^*(2045)\pi $, $ K_2^*(1980)\pi $ and~$K^* b_1$, etc.
 We hope our prediction can be helpful for the future experiment to study this two states and their mixing angle.

\renewcommand{\arraystretch}{1.3}
\begin{table}[htbp]
\caption{The strong decay widths of $K_4(2550)$ assigned to a  $2G^{\prime}$ state, where $s$ and $c$ represent sine and cosine functions. The unit of the width is MeV. \label{tab:k42550}}
\begin{center}
\[\begin{array}{cc}
\hline
\hline
 \text{Decay channel} &\text{Width} \\
 \hline
 \text{Total} & 238 c^2+38c s+218s^2 \\
 K_3^*{(1780)} \pi  & 32.7 c^2+57.5 c s+28.3 s^2 \\
 K \rho _3{(1690)} & 31.2 c^2-55.3 c s+25.7 s^2 \\
 K_4^*{(2045)} \pi  & 25.1 c^2+48.1 c s+24 s^2 \\
K a_2 & 21.8 c^2-35.8 c s+19.8 s^2 \\
 K_2^* \pi  & 23 c^2+34.8 c s+19.4 s^2 \\
  K^*b_1 & 15.1 c^2+25.8 c s+12.3 s^2 \\
K \omega _3{(1670)} & 10.9 c^2-19.3 c s+9.05 s^2 \\
  {K} \rho  & 9.72 c^2-1.24 c s+9.58 s^2 \\
  K^* a_2& 7.94 c^2-7.73 c s+11.9s^2 \\
 K^* a_1 &5.90 c^2+6.60 c s+4.95 s^2 \\
 K_2^* \rho  & 6.29 c^2+5.64 c s+8.34 s^2 \\

 \hline
 \hline
\end{array}\]
\end{center}
\end{table}

\section{Conclusions and discussion}\label{sec5}
In this paper, we have given the analysis of mass spectra of the kaon family via the modified Godfrey-Isgur quark model that includes a color screening effect, and have obtained the structure information of the observed kaon candidates. Then, we have further tested the possible assignments by comparing the theoretical results of their two-body strong decays with the experimental data. Additionally, we have also predicted the behaviors of some partial decay widths of the kaons, which are still missing in experiments. In Table \ref{tab:missingstates}, we summarize the mass and main decay modes of  these  states, by which experiment may carry out the search for them.
\renewcommand{\arraystretch}{1.3}

\begin{table}[htbp]
\caption{The mass and the important  strong decay channels for some predicted kaon states which can be helpful to future search for them in experiments.  The units of the mass and width are MeV. \label{tab:missingstates}}
\begin{center}
\[\begin{array}{cccc}
\hline
\hline
 \text{State} &\text{Assignment}&  \text{Mass}& \text{Main decay channels} \\\midrule[1pt]
 K_1{(2030)}& 2P^\prime&\sim 2030&K^*\pi,{K\rho} \\
 K_2{(1990)} & 2D&\sim 1994& K^*\pi, K\rho \\
 K_3{(2075)}  &1F& \sim 2075&K_3^*(1780)\pi,{K^*\rho} \\
K _3(2360) &2F^\prime&\sim  2362& K_3^*(1780)\pi, K\rho_3(1690) \\
 K_4(2310)  & 1G& \sim 2309&K_4^*(2045) \pi, K_3^*(1780)\pi\\
  K_4(2550)& 2G^\prime&\sim  2550& K_3^*(1780)\pi, K\rho_3(1690) \\
 \hline
 \hline
\end{array}\]
\end{center}
\end{table}
 This study is crucial to establish the kaon family and future search for their higher excitations.
We have discussed the possible assignments to the kaons listed in PDG. The
main task of the present work has been a calculation
of the spectra and OZI-allowed two-body strong decays of the kaon family, which can test the
possible assignments to the kaons. In Sections II and III, we have
discussed these points in detail. The predicted decay
behaviors of the discussed kaons can provide
valuable information for further experimental study in
the future.

At present, experimental information on the kaons
is not abundant. Thus, we suggest
to do more experimental measurements of the resonance
parameters and to search for the missing main decay
channels. Such an effort will be not only helpful to
establish the kaon family in experiments  {but is also valuable to study the production of hidden-charm pentaquarks $P_c(4380)$ and $P_c(4450)$ by analyzing  $\Lambda_b\to J/\psi p K$ \cite{Aaij:2015tga} which has a close relation to the understanding the kaon family}.
With experimental progress, the exploration of the kaons will become a major issue in hadron physics,
provided by good platforms in the BESIII, BelleII, and COMPASS
experiments. We hope that, inspired by this work, more
experimental and theoretical studies of high-spin states are
conducted in the future.

\vfil
\section*{Acknowledgments}
This  work is supported in part by National Natural Science Foundation of China under the
Grant No. 11222547 and No. 11175073, the Fundamental Research Funds for the Central Universities, the High-End Creative Talent Thousand People Plan of Qinghai Province, No. 0042801 and the Applied Basic Research Project of Qinghai Province, No. 2017-ZJ-748.
Xiang Liu is also supported by the National Program for Support of
Top-notch Young Professionals.
\vfil

\bibliographystyle{apsrev4-1}
\bibliography{ref}
\end{document}